\documentclass[conference,compsoc]{IEEEtran}

\IEEEoverridecommandlockouts
\ifCLASSOPTIONcompsoc
  \usepackage[nocompress]{cite}
\else
  \usepackage{cite}
\fi
\ifCLASSINFOpdf
\else
\fi
\usepackage[table]{xcolor}  
\usepackage{tikz}          
\usepackage{amsmath}
\usepackage[ruled]{algorithm2e}

\usepackage{graphicx}
\usepackage{booktabs}
\usepackage{latexsym}   
\usepackage{array}      
\usepackage{multirow}
\usepackage{subcaption}
\usepackage[noend]{algpseudocode}
\usepackage{threeparttable}
\usepackage{paralist}   
\usepackage{xspace}
\usepackage{color}
\usepackage{adjustbox}
\usepackage{balance}
\usepackage{wasysym}
\usepackage{rotating}   
\usepackage{listings}
\usepackage[T1]{fontenc}  
\usepackage{tabu}
\usepackage{pifont}
\usepackage{framed}

\usepackage[shortlabels]{enumitem}
\setenumerate[1]{itemsep=0pt,partopsep=0pt,parsep=\parskip,topsep=2pt}
\setitemize[1]{itemsep=0pt,partopsep=0pt,parsep=\parskip,topsep=2pt}
\setdescription{itemsep=0pt,partopsep=0pt,parsep=\parskip,topsep=2pt}
\usepackage{multirow}
\usepackage{url}            

\usepackage[normalem]{ulem}

\usepackage{threeparttable}
\usepackage{pifont}
\usepackage{tikz,float}
\usepackage{tkz-euclide}
\usepackage{tabularx}

\usepackage{xpatch}
\usepackage[frozencache,cachedir=.]{minted}

\usepackage{ragged2e}

\usepackage{amssymb}
\usepackage{cite}
\usepackage[hidelinks]{hyperref}
\usepackage{cleveref}

\newcommand{\blue}[1]{\textcolor[rgb]{0.00,0.00,1.00}{#1}}

\definecolor{wheat1}{rgb}{1.000000,0.905882,0.729412}
\definecolor{LightGray}{rgb}{0.827451,0.827451,0.827451}

\newcolumntype{a}{>{\columncolor{wheat1}}l}

\definecolor{mygreen}{rgb}{0,0.6,0}
\definecolor{mygray}{rgb}{0.5,0.5,0.5}
\definecolor{mymauve}{rgb}{0.58,0,0.82}
\definecolor{darkblue}{rgb}{0.0,0.0,0.6}
\definecolor{maroon}{RGB}{102, 0, 0}
\definecolor{Maroon}{cmyk}{0,0.87,0.68,0.32}
\definecolor{darkred}{RGB}{139, 0, 0}
\definecolor{forestgreen}{RGB}{34, 139, 34}

\lstdefinelanguage{XML}
{
  basicstyle=\ttfamily\small,   
  morestring=[b]",
  moredelim=[s][\color{darkblue}]{<}{\ },
  moredelim=[s][\color{darkblue}]{</}{>},
  moredelim=[l][\color{darkblue}]{/>},
  moredelim=[l][\color{darkblue}]{>},
  morecomment=[s]{<?}{?>},
  morecomment=[s]{<!--}{-->},
  stringstyle=\color{darkred},
  identifierstyle=\color{mymauve}
}

\lstdefinestyle{customJava}{
  breaklines=true,
  keepspaces=true,
  frame=single,
  language=Java,
  showstringspaces=false,
  basicstyle=\footnotesize\ttfamily,
  keywordstyle=\color{blue},
  otherkeywords={+, getIntent},
  numbers=left,
  numbersep=5pt,
  numberstyle=\scriptsize\color{black},
  rulecolor=\color{black},
  stepnumber=1,
  tabsize=2,
  commentstyle=\itshape\color{green!40!black},
  stringstyle=\color{orange},
  emph=[1]  
  {
        do,
        try,
        new,
        catch,
        while,
        SecProvider,
        SecReceiver,
        SecService,
        SecActivity,
        SecSink,
  },
  emphstyle=[1]{\color{darkred}},
  emph=[2]  
  {
        @Override,
  },
  emphstyle=[2]{\color{purple!40!black}},
  belowskip=-1em, 
}

\newif\ifANNOYMIZE
\ANNOYMIZEtrue

\newif\ifACM
\ACMfalse 

\ifACM
\fi

\ifACM

\else

\fi

\ifACM

\else

\fi

\newcommand{\name}{FinDet\xspace}
\newcommand{\baselinename}{Baseline\textsubscript{NL}}\xspace
\newsavebox{\bigimage} 

\usepackage{array}

\newcolumntype{L}[1]{>{\raggedright\let\newline\\\arraybackslash\hspace{0pt}}m{#1}}
\newcolumntype{C}[1]{>{\centering\let\newline\\\arraybackslash\hspace{0pt}}m{#1}}
\newcolumntype{R}[1]{>{\raggedleft\let\newline\\\arraybackslash\hspace{0pt}}m{#1}}

\makeatletter
\chardef\TPT@@@asteriskcatcode=\catcode`*
\catcode`*=11
\xpatchcmd{\threeparttable}
  {\TPT@hookin{tabular}}
  {\TPT@hookin{tabular}\TPT@hookin{tabu}}
  {}{}
\catcode`*=\TPT@@@asteriskcatcode
\makeatother
\usepackage{tcolorbox}
\tcbuselibrary{breakable, skins}
\tcbset{%
  label begin/.style={label={#1}},
  label end/.style={after upper=\label{#1}}
}
\newtcolorbox[%
auto counter]{mybox}[2][]{%
  enhanced jigsaw,
  breakable,
  #1}

\definecolor{verylightgray}{rgb}{.97,.97,.97}
\definecolor{codegreen}{rgb}{0,0.55,0}

\lstdefinelanguage{Solidity}{
	keywords=[1]{anonymous, assembly, assert, balance, break, call, callcode, case, catch, class, constant, continue, constructor, contract, debugger, default, delegatecall, delete, do, else, emit, event, experimental, export, external, false, finally, for, function, gas, if, implements, import, in, indexed, instanceof, interface, internal, is, length, library, log0, log1, log2, log3, log4, memory, modifier, new, payable, pragma, private, protected, public, pure, push, require, return, returns, revert, selfdestruct, send, solidity, storage, struct, suicide, super, switch, then, this, throw, transfer, true, try, typeof, using, value, view, while, with, addmod, ecrecover, keccak256, mulmod, ripemd160, sha256, sha3}, 
	keywordstyle=[1]\color{blue}\bfseries,
	keywords=[2]{address, bool, byte, bytes, bytes1, bytes2, bytes3, bytes4, bytes5, bytes6, bytes7, bytes8, bytes9, bytes10, bytes11, bytes12, bytes13, bytes14, bytes15, bytes16, bytes17, bytes18, bytes19, bytes20, bytes21, bytes22, bytes23, bytes24, bytes25, bytes26, bytes27, bytes28, bytes29, bytes30, bytes31, bytes32, enum, int, int8, int16, int24, int32, int40, int48, int56, int64, int72, int80, int88, int96, int104, int112, int120, int128, int136, int144, int152, int160, int168, int176, int184, int192, int200, int208, int216, int224, int232, int240, int248, int256, mapping, string, uint, uint8, uint16, uint24, uint32, uint40, uint48, uint56, uint64, uint72, uint80, uint88, uint96, uint104, uint112, uint120, uint128, uint136, uint144, uint152, uint160, uint168, uint176, uint184, uint192, uint200, uint208, uint216, uint224, uint232, uint240, uint248, uint256, var, void, ether, finney, szabo, wei, days, hours, minutes, seconds, weeks, years},	
	keywordstyle=[2]\color{teal}\bfseries,
	keywords=[3]{block, blockhash, coinbase, difficulty, gaslimit, number, timestamp, msg, data, gas, sender, sig, value, now, tx, gasprice, origin},	
	keywordstyle=[3]\color{violet}\bfseries,
	identifierstyle=\color{black},
	sensitive=false,
	comment=[l]{//},
	morecomment=[s]{/*}{*/},
	commentstyle=\color{gray}\ttfamily,
        stringstyle=\color{blue}\ttfamily,
	morestring=[b]',
	morestring=[b]"
}

\lstdefinelanguage{json}{
  basicstyle=\scriptsize\ttfamily,
  numbers=left,
  numberstyle=\scriptsize\color{gray},
  numbersep=6pt,
  tabsize=2,
  breaklines=true,
  showstringspaces=false,
  frame=none,
  literate=
    *{0}{{{\color{numb}0}}}{1}
     {1}{{{\color{numb}1}}}{1}
     {2}{{{\color{numb}2}}}{1}
     {3}{{{\color{numb}3}}}{1}
     {4}{{{\color{numb}4}}}{1}
     {5}{{{\color{numb}5}}}{1}
     {6}{{{\color{numb}6}}}{1}
     {7}{{{\color{numb}7}}}{1}
     {8}{{{\color{numb}8}}}{1}
     {9}{{{\color{numb}9}}}{1}
    {:}{{{\color{punct}{:}}}}{1}
    {,}{{{\color{punct}{,}}}}{1}
    {\{}{{{\color{delim}{\{}}}}{1}
    {\}}{{{\color{delim}{\}}}}}{1}
    {[}{{{\color{delim}{[}}}}{1}
    {]}{{{\color{delim}{]}}}}{1},
  morestring=[b]",
  stringstyle=\color{string},
  commentstyle=\color{comment},
  keywordstyle=\color{keyword},
}

\usepackage{xcolor}
\usepackage{xcolor}
\definecolor{keyword}{rgb}{0,0,0.6}   
\definecolor{string}{rgb}{0.5,0,0}   
\definecolor{numb}{rgb}{0,0,0}       
\definecolor{punct}{rgb}{0,0,0}      
\definecolor{delim}{rgb}{0,0,0}      
\definecolor{comment}{rgb}{0,0,0}    

\tcbuselibrary{listingsutf8}

\newtcolorbox{ResultBox}{
  listing only,
  listing options={
    basicstyle=\ttfamily\scriptsize,  
    breaklines=true,
    columns=fullflexible,
  },
  colback=gray!10,
  colframe=black,
  left=1mm,
  right=1mm,
  top=1mm,
  bottom=1mm,
  boxrule=0.5pt,
  fontupper=\scriptsize\ttfamily,  
}

\tcbset{
  answerbox/.style={
    colback=blue!5!white,
    colframe=blue!75!black,
    fonttitle=\bfseries,
    title=Answer,
    boxrule=0.5pt,
    arc=1mm,
    left=1mm,
    right=1mm,
    top=1mm,
    bottom=1mm,
    opacityfill=0.05,
    breakable,
  }
}

\tikzset{%
	pics/sema/.style args={#1/#2/#3}{code={%
			\ifstrequal{#2}{0}{%
				\draw[fill=#1] (0,0) circle (0.34em){};
			}{%
				\tkzDefPoint(0,0){O}
				\tkzDrawSector[R,fill=#3](O,0.34em)(90,90-#2)
				\tkzDrawSector[R,fill=#1](O,0.34em)(90-#2,90-360)
			}
	}},
}
\begin{document}
%


\title{Revealing Adversarial Smart Contracts through Semantic Interpretation and Uncertainty Estimation}


\author{
\IEEEauthorblockN{Yating Liu, Xing Su, Hao Wu, Sijin Li, Yuxi Cheng, Fengyuan Xu$^*$\thanks{$^*$Fengyuan Xu is the corresponding author.}, Sheng Zhong}

\IEEEauthorblockA{National Key Lab for Novel Software Technology, Nanjing University, Nanjing, Jiangsu, China \\
Emails: \{yatingliu, xingsu, sijinli, yuxicheng\}@smail.nju.edu.cn, \{hao.wu, fengyuan.xu, zhongsheng\}@nju.edu.cn}
}


%


\maketitle

\begin{abstract}
    Adversarial smart contracts, mostly on EVM-compatible chains like Ethereum and BSC, are deployed as EVM bytecode to exploit vulnerable smart contracts for financial gain. Detecting such malicious contracts at the time of deployment is an important proactive strategy to prevent losses from victim contracts. It offers a better cost-benefit ratio than detecting vulnerabilities on diverse potential victims. However, existing works are not generic with limited detection types and effectiveness due to imbalanced samples, while the emerging LLM technologies, which show their potential in generalization, have two key problems impeding its application in this task: hard digestion of compiled-code inputs, especially those with task-specific logic, and hard assessment of LLM's certainty in its binary (yes-or-no) answers. Therefore, we propose a generic adversarial smart contracts detection framework \name, which leverages LLM with two enhancements addressing the above two problems. \name takes as input only the EVM bytecode contracts and identifies adversarial ones among them with high balanced accuracy. The first enhancement extracts concise semantic intentions and high-level behavioral logic from the low-level bytecode inputs, unleashing the LLM reasoning capability restricted by the task input. The second enhancement probes and measures the LLM uncertainty to its multi-round answering to the same query, improving the LLM answering robustness for binary classifications required by the task output. Our comprehensive evaluation shows that \name achieves a BAC of 0.9374 and a TPR of 0.9231, significantly outperforming existing baselines. It remains robust under challenging conditions including unseen attack patterns, low-data settings, and feature obfuscation. \name detects all 5 public and 20+ unreported adversarial contracts in a 10-day real-world test, confirmed manually.
\end{abstract}


%
\IEEEpeerreviewmaketitle

\pagestyle{plain}  

\section{Introduction}
\label{sec:intro}




While blockchain technology is advancing rapidly, its decentralized and anonymous nature also creates opportunities for malicious activities. In 2024, security incidents led to losses exceeding \$2.01 billion~\cite{slowmist2024hacked}. To protect the ecosystem, prior research mainly focused on detecting vulnerable smart contracts. However, many reported vulnerabilities are ultimately unexploitable~\cite{perez2021smart}. Moreover, defenders face an inherent asymmetry, as they must secure against all threats, whereas attackers need only a single successful exploit. Other approaches monitor adversarial transactions in public mempools, but they cannot observe transactions in private pools and can only analyze malicious behavior after it has been broadcast or included on-chain.





\begin{figure}[t!]
    \includegraphics[width=\linewidth]{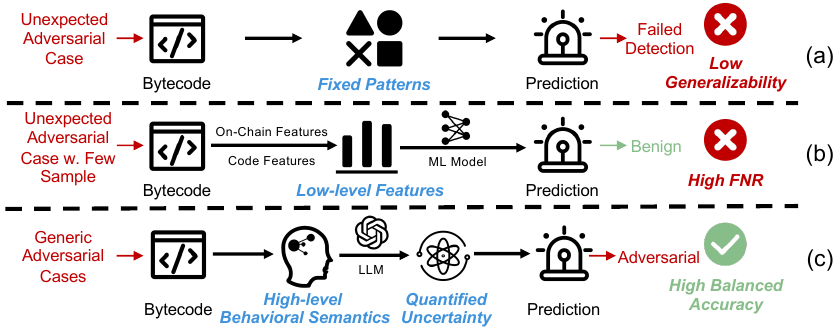}
    \caption{Comparison of adversarial contract detection methods: 
    (a) Rule-based methods rely on fixed patterns and fail on unexpected attacks. 
    (b) ML-based methods leverage low-level features but struggle with few-shot or unseen cases, resulting in high FNR.
    (c) \name employs semantic reasoning for robust adversarial detection.}
    \label{fig:comp_overall_cost_and_token_dist}
\end{figure}

To address these limitations, recent research efforts have shifted toward adversarial contract detection. Adversarial contracts are deliberately crafted smart contracts that exploit known vulnerabilities or manipulate protocol logic for illicit gain. These approaches focus on contracts that are both exploitable and likely to be deployed in real attacks. By identifying these high-impact threats ahead of execution, defenders can intervene before damage occurs and significantly reduce the burden of manual auditing.

Most adversarial contracts exist only as bytecode to conceal their malicious intent. As shown in \figurename~\ref{fig:comp_overall_cost_and_token_dist}, rule-based methods~\cite{yang2024uncover,zhang2025following} detect adversarial behaviors by disassembling bytecode and extracting control and data flows, which are then analyzed using manually defined heuristics and rule-based logic. However, these methods are effective only for single attack types and lack the flexibility to generalize beyond predefined patterns.
To enable detection of multiple attack types, Machine learning (ML)-based methods~\cite{ren2025lookahead,wang2024skyeye} leverage low-level intermediate representation (IR) and on-chain features to train detection models. However, these methods suffer from high false negative rates, reaching 35.16\% for Lookahead~\cite{ren2025lookahead} and 18.68\% for Skyeye~\cite{wang2024skyeye}. Moreover, they face two fundamental limitations: limited generalization ability, which impedes the detection of unseen attack patterns as adversarial behaviors continuously evolve beyond the training data, and a strong dependence on large amounts of labeled data, which is often scarce and costly, resulting in significant performance degradation when training samples are insufficient.

\begin{figure}[htbp]
    \centering
    \begin{minipage}[t]{0.49\linewidth}
        \centering
        \includegraphics[width=\linewidth]{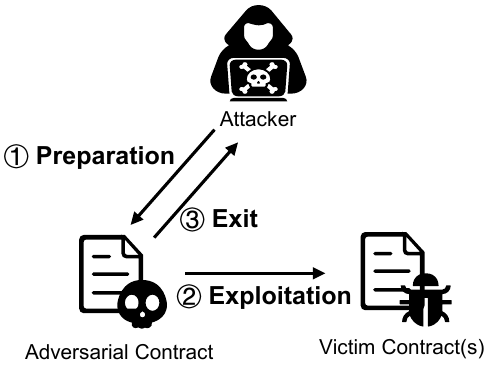}
        \caption{Three-phase adversarial lifecycle.}
        \label{fig:motivation_3_phase}
    \end{minipage}
    \hfill
    \begin{minipage}[t]{0.49\linewidth}
        \centering
        \includegraphics[width=\linewidth]{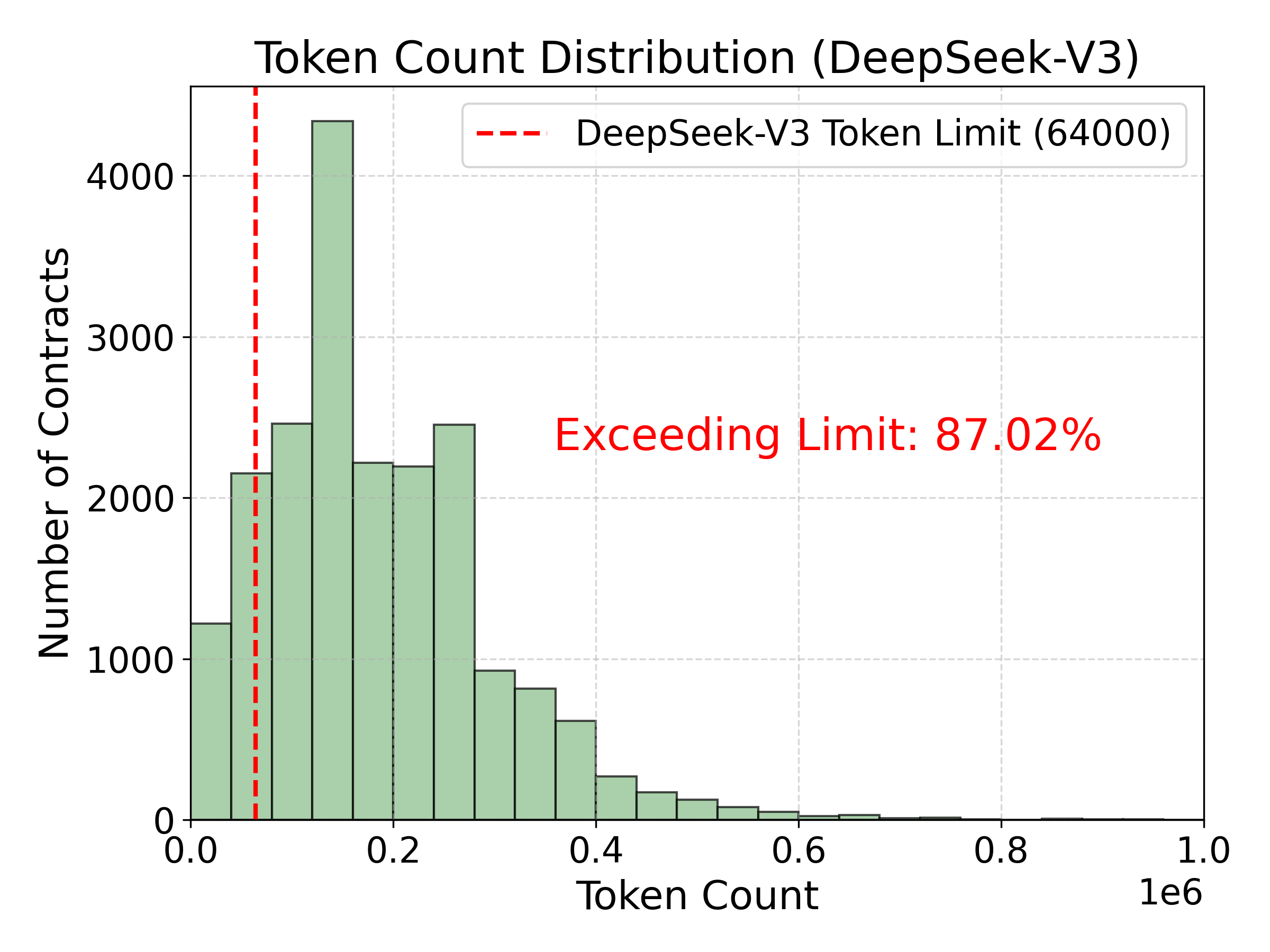}
        \caption{Token distribution of TAC on DeepSeek-V3.}

        \label{fig:elipmoc_token_count}
    \end{minipage}
\end{figure}




Encouragingly, large language models (LLMs) have demonstrated tremendous potential in understanding advanced code representations (e.g., source code), exhibiting strong capabilities in capturing behavioral semantics and interpreting smart contract logic. These strengths make LLMs a promising alternative for detecting adversarial smart contracts, especially in scenarios where labeled training data is scarce and costly to obtain. 
However, this approach faces several key challenges:

\begin{itemize}

    \item \textbf{C1: Limited Understanding of Low-Level Intermediate Representations.} LLMs are primarily pre-trained on high-level source code, and thus lack the capability to accurately understand the low-level semantics of EVM bytecode. One alternative approach is to lift bytecode into IR, such as using Elipmoc~\cite{grech2022elipmoc} to transform bytecode into three-address code (TAC) for improved interpretability.  However, as shown in \figurename~\ref{fig:elipmoc_token_count}, 87.02\% of TAC representations exceed typical LLM input limits, leading to frequent detection failures. Fine-tuning LLMs directly on bytecode is possible but costly and limited by scarce labeled data.

    \item \textbf{C2: Diverse and Evolving Attack Strategies.} Adversarial contracts employ a wide range of attack methods that continuously evolve over time. Relying solely on superficial features of previously known attacks for decision-making limits the ability to detect unseen and emerging attack types effectively.
    
    \item \textbf{C3: Hallucination and Challenging Uncertainty Assessment.} LLMs tend to lose focus when processing complex logic, often resulting in high false positives due to conservative decisions under uncertainty~\cite{sun2024gptscan}; since most mainstream LLMs are closed-source and operate as black-box systems, detecting and quantifying their hallucinations is particularly challenging and reduces detection reliability.
\end{itemize}

In this work, to leverage the capabilities of LLMs while addressing the above challenges, we propose \name, a generic adversarial smart contract detection method that enhances high-level behavioral semantics and enables quantified assessment of LLM uncertainty.
\name relies solely on bytecode thus enabling early detection during the pre-deployment phase, and is also capable of robustly identifying previously unseen attack patterns.
It operates in two stages.  
In Stage~I, the bytecode is lifted into a semi-structured natural language (NL) description, based on which we perform general-purpose analysis from the perspective of the contract’s overall semantics, and attack-specific analysis grounded in the semantics of operational logic.  
In Stage~II, we conduct fine-grained uncertainty probing and assessment. By leveraging the multi-view semantic information obtained in Stage~I through targeted questions and performing reliable fusion, \name achieves robust detection results.

To address \textbf{C1}
, \name adopts a bytecode lifting mechanism that translates raw bytecode into semi-structured NL descriptions. This process preserves critical semantic information while enhancing interpretability. Unlike traditional intermediate-representation-level static analyses that struggle to capture subtle contract logic, this semantic elevation enables more accurate and robust detection of adversarial behaviors.
\textbf{C2} stems from the complexity and rapid evolution of adversarial attack strategies, making generic detection difficult. To tackle this, we analyzed many adversarial cases and identified a fundamental three-stage operational pattern for attacker-deployed adversarial contracts aimed at profit extraction: \ding{172}~\textit{\textbf{Preparation}}, \ding{173}~\textit{\textbf{Exploitation}}, and \ding{174}~\textit{\textbf{Exit}} (see \figurename~\ref{fig:motivation_3_phase} and \textbf{Obs2} in \S\ref{sec:observations}). \name systematically analyzes fund-flow reachability across these stages, capturing core operational logic that differentiates such adversarial contracts from benign ones.
As to \textbf{C3}, we first transform the yes-or-no detection query posed to LLMs into a task of assigning probabilities across four levels of uncertainty. We then repeat the same task with disturbance prompts to assess how confident the LLMs are. The uncertainty in their responses is quantified by the entropy of the resulting probability distributions~\cite{shannon1948mathematical}, which is further used to derive the final yes-or-no answer with high confidence.


In our experiments, \name achieved a state-of-the-art balanced accuracy (BAC) of 0.9374, surpassing the baselines Lookahead and Skyeye by 13.83\% and 3.59\%, respectively. It also attained a recall of 0.9231 on the adversarial class, outperforming Lookahead (42.37\%) and Skyeye (13.51\%), while maintaining an average cost of \$0.003 per contract.
Notably, \name consistently demonstrated robust performance under challenging scenarios, including unseen attack patterns, insufficient training data, and on-chain feature obfuscation, where baseline methods suffered significant performance degradation.
During a 10-day real-world detection, \name successfully identified 29 adversarial contracts, among which 5 were previously undisclosed and exhibited clear adversarial intent.
To support systematic evaluation and future research, we further curated the first and the largest dataset of adversarial contracts with normalized categories, including 455 adversarial and 20,000 benign samples, of which 200 adversarial contracts are labeled with normalized attack types.

We summarize our contributions as follows:
\begin{itemize}
    \item We propose \name, the first generic and training-free detection framework that leverages the semantic understanding capabilities of LLMs to identify adversarial smart contracts directly from EVM bytecode, enabling detection in the pre-deployment phase. \name demonstrates robust performance under challenging scenarios such as unseen attack patterns and insufficient training data, and can seamlessly work with different LLMs without requiring fine-tuning.

    \item We enhance semantic understanding by integrating fund-flow reachability analysis and multi-view semantic reasoning to capture fundamental adversarial patterns, both based on the lifted semi-structured NL descriptions derived from bytecode.
    \item We design a novel quantitative uncertainty assessment of LLM outputs by extending labels to a fine-grained scale, incorporating multiple probing prompts and applying entropy-based fusion to produce reliable uncertainty estimates.
    \item \name achieved state-of-the-art performance with a BAC of 0.9374 (surpassing the baselines by 13.83\% and 3.59\%) and an adversarial recall (i.e., TPR) of 0.9231 (exceeding the baselines by 42.37\% and 13.51\%). Furthermore, \name identified 29 previously undisclosed adversarial contracts in real-world.
\end{itemize}


\section{Background}
\label{sec:background}

\begin{figure*}[t!]
    \includegraphics[width=\linewidth]{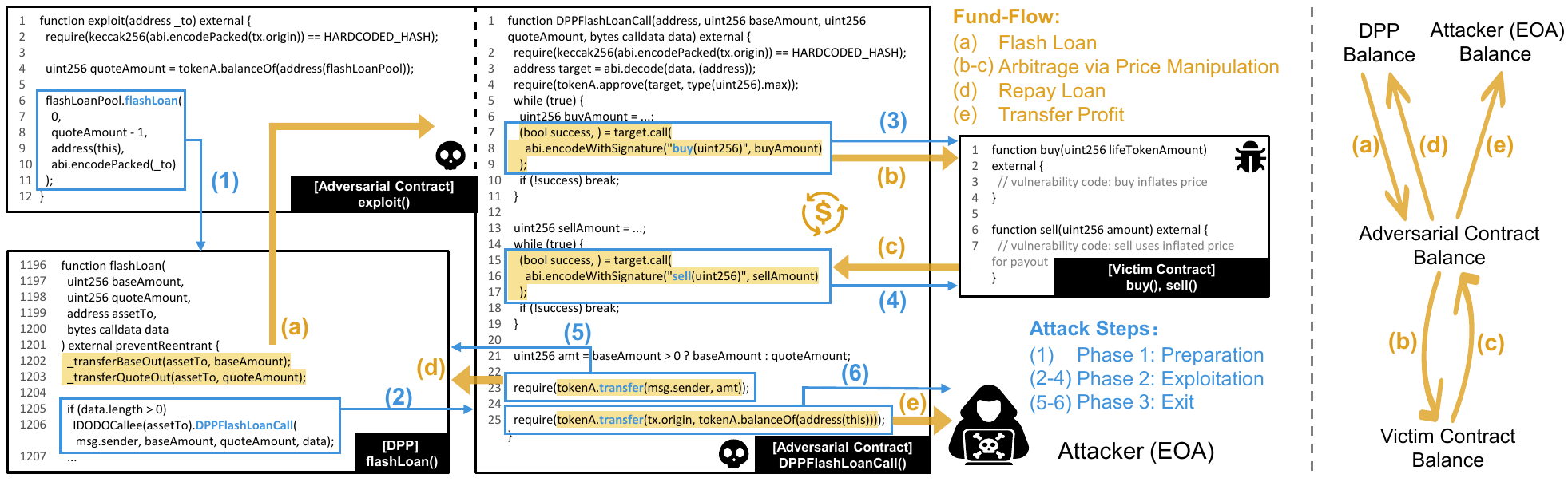}
    \caption{
     Decompiled Solidity snippet of an adversarial contract (see \href{https://bscscan.com/address/0x75f2002937507b826b727170728595fd45151d0f}{BscScan link}) involved in a flash loan attack on April 26, 2025.
    This case illustrates a typical flash loan exploit, following the three-phase fund-flow pattern.  
    The right part visualizes the corresponding fund transfers across different addresses during the attack.
    }
    \label{fig:case_study_code}
\end{figure*}





\subsection{Decentralized Finance}

Blockchain provides a decentralized, peer-to-peer, and verifiable infrastructure that records and secures transactions through distributed consensus while preserving user anonymity. 
Built atop this foundation, Decentralized Finance (DeFi) enables permissionless and composable financial services through smart contracts. 
Its core elements are accounts and transactions, which define asset ownership and drive contract execution.

\textbf{Accounts.} Ethereum supports two account types: Externally Owned Accounts (EOAs) controlled by private keys, and smart contract accounts governed by deployed code.

\textbf{Transactions.} Transactions underpin DeFi execution. EOAs initiate external transactions, while smart contracts generate internal transactions to implement protocol logic.

\subsection{Vulnerabilities and Adversarial Contracts}


Smart contracts are autonomous programs that manage assets and enforce logic in DeFi protocols, featuring immutability, composability, and permissionless interaction. 
Vulnerable contracts~\cite{sendner2024large} contain unintended flaws that can be exploited to cause incorrect behavior or asset loss, such as reentrancy or price manipulation.
In contrast, adversarial contracts~\cite{yang2024uncover} are deliberately crafted malicious contracts designed to exploit known vulnerabilities or manipulate protocol logic for illicit gain.

\subsection{Large Language Models for Blockchain Security}


The emergence of LLMs marks a major advance in language understanding and generation. Built on Transformer architectures with billions of parameters, they exhibit strong reasoning and semantic comprehension. In blockchain, LLMs have gained attention for tasks like smart contract auditing~\cite{sun2024gptscan,liu2024propertygpt}, on-chain transaction analysis~\cite{hu2024zipzap}, and dynamic contract analysis~\cite{sun2025adversarial}. These applications highlight LLMs’ potential as powerful tools to enhance blockchain security.

\section{Threat Model, Motivation and Observations}
\label{sec:observations_and_threat_model}

\subsection{Threat Model, Scope, and Assumptions}

\name targets adversarial contracts crafted to exploit on-chain vulnerabilities: attackers deploy such contracts to interact with vulnerable victim contracts, exploit their flaws, and ultimately extract illicit proceeds. We assume attackers have full access to public information including victim contract source or decompiled code and can deploy arbitrary contracts, submit transactions, and leverage private submission channels (e.g., Flashbots) to conceal their activity.

Attacks unrelated to contract-level vulnerabilities (e.g., private-key leaks or off-chain compromises) and those carried out by privileged users (e.g., rug pulls) are out of scope. We also exclude attacks executed solely via crafted transactions without a deployed adversarial contract.

Detection occurs pre-deployment (mempool stage), relying only on bytecode without runtime blockchain data. Our approach generalizes across variants by analyzing semantic intent rather than relying on brittle signatures or handcrafted datasets. A detection is considered successful if malicious behavioral indicators are identified prior to deployment, allowing the contract to be correctly classified as adversarial.

\subsection{Motivation Example}
\label{sec:motivation_example}
\figurename~\ref{fig:case_study_code} illustrates a typical scenario of an attack via flashloans.

The blue arrows indicate the attack steps: 
(1) The attacker calls the \texttt{exploit} function, obtaining a flash loan from the DPP contract to secure initial capital; 
(2) A callback is triggered during the loan, invoking \texttt{DPPFlashLoanCall}, a function controlled by the attacker; 
(3-4) The attacker performs repeated token buy/sell operations to exploit price discrepancies; 
(5) The borrowed funds are then repaid to the lending protocol, and 
(6) the extracted profit is transferred to the attacker.

The yellow arrows illustrate the fund-flow throughout the attack:
(a) a flash loan is issued from the DPP contract to the adversarial contract;
(b) the adversary initiates a buy operation with a small input to artificially inflate the token price;
(c) this is followed by a sell operation to extract profit based on the manipulated price;
(d) the borrowed funds are then repaid to the DPP contract; and finally,
(e) the remaining profit is transferred from the adversarial contract to the attacker’s address.
This motivating example reveals typical behavioral patterns of adversarial contracts. We elaborate on these observations in the following section.

\subsection{Observations}
\label{sec:observations}

After identifying the core challenges, we further make the following key observations from our empirical analysis, which directly inform the design of our system.

\textbf{Obs1: Long-Form Inputs Cause Loss of Contextual Focus.}
\label{obs:loss_focus}
We find that LLMs struggle to maintain semantic coherence when presented with entire contracts as monolithic inputs. Instead of capturing the overarching behavioral intent, the model often fixates on isolated or syntactically unusual fragments, leading to erroneous reasoning. For example, as shown in \figurename~\ref{fig:ablation_case_prompt4} in Appendix~\ref{sec:ablation_case}, a benign contract was misclassified due to its intricate logic structure and dense permission checks. This observation motivates our use of structured prompting and hierarchical summarization to preserve contextual clarity and guide the model’s attention more effectively.

\textbf{Obs2: Adversarial Fund-Flows Capture the Fundamental Three-Stage Attack Lifecycle.}
\label{obs:fund_flow}
From empirical analysis of adversarial contracts, we observe that, within our scope (i.e., attacks where an attacker deploys a contract to exploit a victim contract for financial gain), the driving force is illicit fund acquisition, and the behavior can be abstracted into three essential stages (see \figurename~\ref{fig:motivation_3_phase}) that are fundamentally required to achieve the attacker's goal:  \ding{172}~\textit{\textbf{Preparation}}, where the attacker first deploys the adversarial contract and completes necessary setup (e.g., obtaining authorizations, funding, or initializing state); \ding{173}~\textit{\textbf{Exploitation}}, where the contract executes logic to leverage the victim's vulnerabilities; and \ding{174}~\textit{\textbf{Exit}}, where, since adversarial contracts are often deployed for one-off attacks, the illicit proceeds are typically transferred to attacker-controlled addresses .

We emphasize that these stages do not constitute a general abstraction of all blockchain attacks, but instead represent the necessary elements to achieve financial gain within the threat model of attacker-deployed adversarial contracts. Under this abstraction, fund-flow analysis, tracing incoming funds, internal transfers, and outgoing flows, naturally captures the chain-of-attack reasoning required to distinguish adversarial contracts from benign ones.

\textbf{Obs3: Ambiguous Patterns Trigger Conservative Reasoning.}
\label{obs:conservative_resoning}
When encountering gray-area behaviors such as redundant access checks, complex fallback logic, or reentrant fund management, LLMs tend to default to conservative classifications due to uncertainty, often leading to false positives. These behaviors may be essential for security or robustness in benign contracts but are misinterpreted as obfuscation or adversarial intent. This insight underscores the need to quantify model uncertainty, leading to our entropy-based scoring and fusion mechanism, which interprets model outputs as graded confidence signals rather than binary decisions.

\section{Overview of \name}
\label{sec:overview}



\begin{figure}[t!]
    \includegraphics[width=\linewidth]{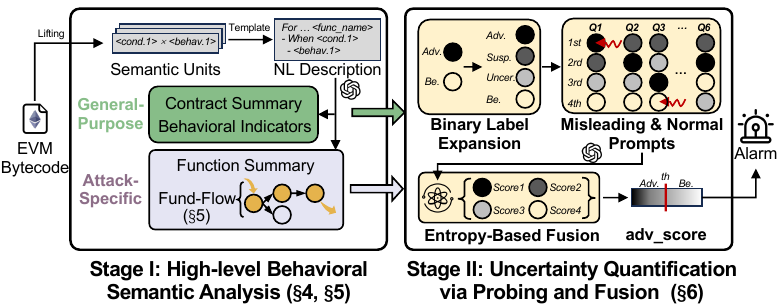}
    \caption{A high-level workflow of \name.}
    \label{fig:overview}
\end{figure}

This section presents the overall design of \name, a general framework that leverages the strong semantic understanding capabilities of LLMs to detect adversarial smart contracts directly from EVM bytecode. \name supports generic detection and is capable of identifying previously unseen attack patterns. As illustrated in \figurename~\ref{fig:overview}, the framework consists of two main stages: high-level behavioral semantic analysis (Stage~I) and uncertainty quantification via probing and fusion (Stage~II). We parallelize LLM queries within Stage I and Stage II separately to improve efficiency.

In Stage~I, we begin by lifting raw EVM bytecode into semantic units and transforming them into semi-structured NL descriptions through templates (\S\ref{sec:lifting}) that apply tailored grammatical rules to express different semantic unit types, addressing the limited understanding of low-level intermediate representations (\textbf{C1}).
We perform general-purpose analysis (\S\ref{sec:attack_intent}) to understand the overall purpose of the contract and extract sensitive semantics. We also conduct attack-specific analysis (\S\ref{sec:attack_action}) by examining each function individually to summarize its intent, suspicious behaviors, and supporting evidence.
In addition, a key component of this analysis is the fund-flow reachability analysis (\S\ref{sec:fund_flow}), which targets diverse and evolving attack strategies (\textbf{C2}) based on the three-phase adversarial pattern shown in \figurename~\ref{fig:motivation_3_phase}.
This approach traces attacker-controlled fund-related operations and uncovers potential exploitation paths.

In Stage~II, to address LLM hallucination and the difficulty of assessing uncertainty, especially in black-box scenarios (\textbf{C3}), we introduce a probing- and fusion-based method that extends binary classification into fine-grained risk assessment (\S\ref{sec:uncertainty}). Leveraging normal and misleading probes with entropy-based fusion, we quantify and rank LLM uncertainty to achieve reliable detection.

\subsection{Natural Language Description Generation}
\label{sec:lifting}

As shown in \figurename~\ref{fig:elipmoc_token_count}, converting bytecode into TAC via Elipmoc~\cite{grech2022elipmoc} makes over 87\% of contracts exceed typical LLM token limits, highlighting the need for more compact, high-level inputs.


Smart contracts automatically execute agreements when certain conditions are met and can thus be seen as a set of conditional behaviors, referred to as semantic units, representing behaviors triggered under specific conditions. We lift bytecode into semantically faithful intermediate representations, the semantic units, by disassembling it into a control-flow graph and extracting semantically relevant operations along each execution path.
For each condition and behavior, we employ predefined templates to transform them into semantically equivalent NL descriptions. Condition templates (e.g., ``\texttt{when...}'', ``\texttt{for...}'') express the triggering logic, while behavior templates summarized in \tablename~\ref{table:behavior_types} describe the resulting actions. 

Unlike LLM-based decompilation approaches~\cite{su2025disco, david2025decompiling}, our method adopts a precise and interpretable NL representation that \textbf{does not rely on LLMs} for its generation. We retain the NL representation for three reasons.
First, regenerating code through LLMs may introduce hallucinations and semantic inconsistencies. As our objective is to extract and reason about semantics rather than reproduce the original source code, we instead preserve a faithful NL description.
Second, code generation substantially increases token consumption and inference latency, while low-level alternatives such as TAC remain excessively verbose for long bytecode sequences.
Third, NL provides a concise, high-level abstraction that enables efficient downstream analyses. All subsequent components in our framework directly operate on this representation.

\subsection{General-Purpose Analysis}
\label{sec:attack_intent}

General-purpose analysis strategy begins with understanding the contract's overall intent and core functionality. By capturing this high-level perspective, we can more effectively interpret the contract's design objectives and evaluate its potential security risks. To achieve this, we prompt a LLM to synthesize information across all functions and generate a comprehensive, contract-level summary, as illustrated 
in \texttt{Task: contract\_summary} in \figurename~\ref{fig:stage1_prompt}. An example summary for the contract depicted in \figurename~\ref{fig:case_study_code} is provided in \figurename~\ref{fig:motivation_case_contract}.
Furthermore, we extract several contract-level indicators from the NL descriptions, including the number and proportion of external calls, unknown functions, and bot-related functions, as well as the presence of fund-transfer behaviors (e.g., operations indicative of attempts to drain the contract balance). These metrics contribute to a holistic understanding of the contract's structural characteristics and its potential security implications.

\subsection{Attack-Specific Analysis}
\label{sec:attack_action}


A central component of our attack-specific analysis is the \textbf{fund-flow reachability analysis}, which traces how attacker-controlled inputs propagate toward sensitive or high-risk operations. Given its importance, we provide a dedicated and detailed discussion in \S\ref{sec:fund_flow}. 

Beyond fund-flow reasoning, we further adopt a fine-grained, function-level behavioral analysis that guides the LLM to examine each function's intended purpose, suspicious behaviors, and supporting evidence, as illustrated in \texttt{Task: function\_summary} in \figurename~\ref{fig:stage1_prompt}. This approach reveals subtle signs of malicious logic that may be overlooked in global-level summaries. While benign contracts typically consist of modular functions with well-defined objectives and appropriate access control, adversarial contracts often embed sensitive operations or privilege escalation within ostensibly benign functions. Accurately identifying each function's intent is thus essential for distinguishing malicious behavior from legitimate functionality. Moreover, detecting suspicious patterns, such as unsafe external calls or attacker-controlled inputs, and extracting concrete evidence helps preserve critical indicators.
As shown in \figurename~\ref{fig:motivation_case_func}, this function-centric analysis is based on the contract presented in \figurename~\ref{fig:case_study_code} and guides the LLM to develop a detailed understanding of the contract's behavior.
We also retain the NL descriptions of functions without identified names, as these are more likely to contain malicious logic.


\section{Fund-Flow Reachability Analysis}
\label{sec:fund_flow}



\begin{figure*}[tb]
    \includegraphics[width=\linewidth]{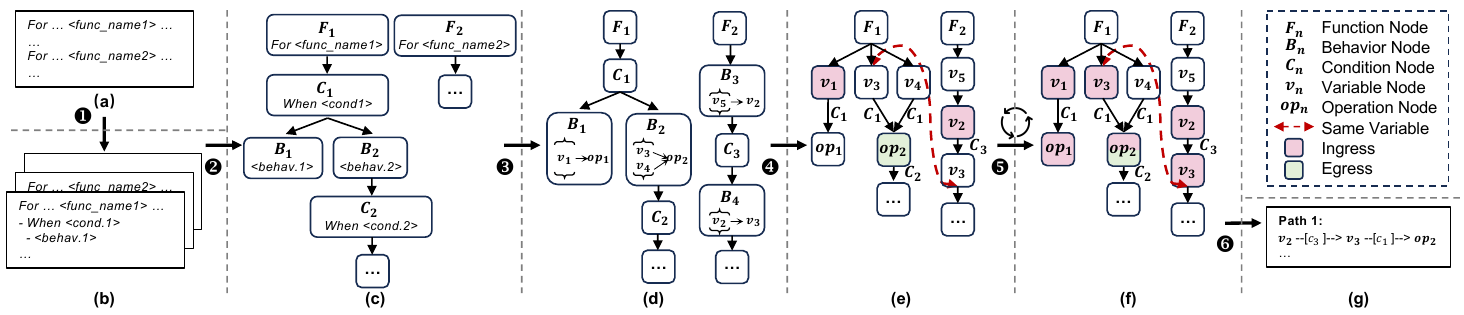}
    \caption{Workflow of fund-flow reachability analysis consisting of six steps: \ding{182} chunking, \ding{183} contract forest construction, \ding{184} behavior node entity resolution, \ding{185} graph transformation \& ingress/egress identification, \ding{186} fund-flow reachability analysis, and \ding{187} reverse pruning \& output.}
    \label{fig:forest_and_taint}
\end{figure*}


Building on the three-stage lifecycle described in \textbf{Obs2}, we focus on fund-flow reachability to capture the ``chain-of-attack'' logic \textbf{inherent to attacker-deployed adversarial contracts}, as part of our analysis in \S\ref{sec:attack_action}. Adversarial contracts are fundamentally driven by illicit financial gain, which leads to deliberate manipulation of fund movements and underscores the necessity of analyzing fund-flows.

At the same time, they differ from benign contracts in how funds are routed: whereas benign contracts enforce transfers through rigid, auditable logic (e.g., owner-only withdrawals or strict conditional checks), adversarial contracts introduce flexible or hidden paths that enable unauthorized movement. These structural differences indicate the feasibility of extracting fund-flow information from the contract's semantic representations, as such manipulations are encoded in the operations within the contract.

We construct a control dependency forest from the NL representation using grammatical rules to get a structured view of the contract and decomposes complex code into manageable structures. We then trace data flow from attacker-controlled ingress to egress that may trigger critical financial operations.

\subsection{Control Dependency Forest Construction}
\label{sec:dependency_forest}

As shown in \figurename~\ref{fig:forest_and_taint} (a-d), we propose a control dependency forest to systematically analyze the semantic structure of smart contracts.  
Given a NL description of a contract (a), we first perform \ding{182} chunking to segment it into a set of function-level descriptions (b).
For each function, we \ding{183} construct a control dependency tree by capturing the dependencies expressed in the description. Specifically, in each tree, every sentence is treated as a node, with the function signature serving as the root.
Each node is heuristically classified into one of four categories:
\begin{itemize}
    \item \textbf{Function:} root node representing a function signature;
    \item \textbf{Behavior:} actions such as \texttt{external calls}, \texttt{assignments}, and \texttt{delegate calls};
    \item \textbf{Condition:} control logic including \texttt{if-else}, \texttt{require}, and loops;
    \item \textbf{Unknown:} fallback nodes for unparsed sentences as a safety mechanism.
\end{itemize}
This classification is feasible because the NL descriptions are generated using a predefined template, ensuring consistent structure and phrasing.
As a result, the entire contract is transformed into a control dependency forest (c).

We further \ding{184} refine the behavior node set by assigning each node a fine-grained behavior type (e.g., \texttt{external call}, \texttt{assignment}) according to its semantic role, as summarized in \tablename~\ref{table:behavior_types}.
For each behavior node \(B_i \in \mathcal{N}_\text{behavior}\), we extract the related entities
including variables $v$ and function operations $op$, and define a propagation tuple:
\begin{align}
    B_i = (\mathcal{X}_\text{src}, x_\text{dst})
    \label{eq:behavior_entity}
\end{align}
where \(\mathcal{X}_\text{src}\) is the set of source entities (inputs) and \(x_\text{dst}\) is the destination entity (output).
We exclude constants such as numeric literals and hardcoded addresses.
To distinguish homonymous entities across functions, we prepend the function name as a namespace to local entities (e.g., \texttt{pancakecall:param1}), while global entities retain their original names (e.g., \texttt{msg.caller}).
Depending on the behavior type, either the source set or destination field in the tuple may be empty.
For the other three node types, we keep their original NL descriptions to preserve full semantic context.
This yields the refined forest of behavior nodes illustrated in \figurename~\ref{fig:forest_and_taint}(d).

\subsection{Graph Transformation}

As illustrated in \figurename~\ref{fig:forest_and_taint} (d-e), we first perform \ding{185} graph transformation to convert the contract forest into a structure suitable for fund-flow reachability analysis. 
For each function tree, we conduct a depth-first traversal from the root node $F_i$, maintaining a set of visited entities with their associated condition sets $(e, \mathcal{C})$, initialized by the formal parameters of $F_i$ and global variables.
When visiting a behavior node $B_i$, we extract its source entities $\mathcal{X}_\text{src}^{(i)}$ and destination entity $x_\text{dst}^{(i)}$. 
If a source $x_j \in \mathcal{X}_\text{src}^{(i)}$ has been visited and $x_\text{dst}^{(i)}$ is new, we connect them with an edge annotated by the combined conditions accumulated along the traversal path:
\[
x_j \xrightarrow{\mathcal{C}} x_\text{dst}^{(i)}
\]
We then mark $x_\text{dst}^{(i)}$ as visited with the corresponding $\mathcal{C}$.

To ensure cross-function consistency, local variables are prefixed with their function names, while global variables retain their original identifiers, as described before. 
Entities sharing the same normalized name across functions are linked to form inter-function edges (e.g., variable \texttt{v3} in \texttt{F1} and \texttt{F2} in \figurename~\ref{fig:forest_and_taint}~(e)). 
Behavior nodes without resolvable entities are discarded.

This process yields a transformed graph where each node (except the root) represents an entity, either a variable or an operation, and all condition nodes are encoded as edge annotations, as shown in \figurename~\ref{fig:forest_and_taint}~(e).

\subsection{Ingress and Egress Identification}

To enable fund-flow reasoning in adversarial contract analysis, we first identify potential entry and exit points of value-related information within contracts, referred to as ingress and egress. 
These anchors, shown in \figurename~\ref{fig:forest_and_taint}~(e-g), enable us to capture how valuable attacker-controlled inputs propagate toward risky operations, forming the foundation for modeling a contract’s fund-flow logic and driving subsequent reachability analysis.
We construct initial sets of ingress and egress reflecting real-world attack semantics, corresponding to the three phases in \figurename~\ref{fig:motivation_3_phase} and detailed in Listing~\ref{lst:predefined_ingress_egress}.

Ingress points represent attacker-controllable or valuable data that may trigger value or permission flows within a contract.
Typical examples include the function caller (\texttt{msg.sender}), the amount of Ether attached to a call (\texttt{msg.value}), the transaction initiator (\texttt{tx.origin}), and the contract’s current balance (\texttt{address(this).balance}). 
We also treat function parameters as contextual ingress points, as they frequently carry user-supplied inputs that affect internal logic and state. 
Together, these constructs capture the primary channels through which external influence can enter a contract’s execution. 
Semantically, they correspond to the \ding{172}~\textit{\textbf{Preparation}} phase of an attack, where adversaries gain capital, trigger flash loans, or manipulate inputs to set up subsequent exploitation.

Egress points, in contrast, denote critical financial or permission-related operations whose execution under attacker influence may lead to fund extraction, unauthorized transfers, or system compromise.
They encompass both standard and low-level behaviors. 
On the standard side, functions such as \texttt{transfer}, \texttt{transferFrom}, and \texttt{approve} are key ERC-20 primitives for asset transfer and allowance control, while \texttt{deposit} and \texttt{withdraw} govern asset custody in DeFi protocols. 
On the low-level side, \texttt{delegatecall} enables arbitrary code execution, and \texttt{selfdestruct} can irreversibly drain contract balances. 
These operations align with the \ding{173}~\textit{\textbf{Exploitation}} phase, where internal states are manipulated for gain, and the \ding{174}~\textit{\textbf{Exit}} phase, where profits are withdrawn or laundered. 
By tracing propagation paths from ingress to egress, we capture the semantic essence of adversarial fund-flows and ensure that our definition of these anchors is grounded in real-world attack semantics rather than arbitrary heuristics.

By matching behavior-node variables against these predefined patterns, we extract ingress and egress node sets $\mathcal{S}_\mathrm{ingress}$ and $\mathcal{S}_\mathrm{egress}$ for each function, serving as semantic anchors for downstream fund-flow and intent analysis.

\begin{lstlisting}[language=Python, caption={Initial ingress and egress for fund-flow-oriented adversarial analysis}, label={lst:predefined_ingress_egress}]
INITIAL_INGRESS = {
    "msg.sender",            # function caller
    "msg.value",             # incoming ETH value
    "tx.origin",             # transaction origin
    "address(this).balance", # contract's balance
    "<function parameters>"  # function parameters
}
INITIAL_EGRESS = {
    "transfer",            # token/native transfer
    "transferFrom",        # delegated transfer
    "approve",             # grant allowance
    "deposit",             # asset deposit
    "withdraw",            # asset withdrawal
    "flashLoan",           # flash loan
    "swapExactTokensForTokens",  # token swap
    "selfdestruct",        # destroy & transfer
    "delegatecall"         # external logic call
}
\end{lstlisting}

\subsection{Reachability Analysis}
\label{sec:reachability}

Given a set of ingress points $\mathcal{S}_{\mathrm{ingress}}$, we perform fund-flow propagation using a depth-first search (DFS) strategy, as outlined in Algorithm~\ref{alg:fund_flow_reachability}. 
Starting from each ingress node, the traversal recursively explores downstream variables; each newly encountered variable is added to the reachable set until no new nodes appear. 
Inter-function propagation is enabled by the normalized naming scheme, which identifies equivalent entities across functions and connects them through inter-function edges. 
To avoid infinite recursion from cycles, a visited set is maintained throughout the process.

After forward propagation, a backward traversal is conducted from the egress points $\mathcal{S}_{\mathrm{egress}}$ to retain only semantically valid paths that originate from an ingress and terminate at an egress, pruning irrelevant branches. 
For example, in \figurename~\ref{fig:forest_and_taint}~(f), the path from $v_1$ to $op_1$ is removed since $op_1 \notin \mathcal{S}_{\mathrm{egress}}$, whereas $op_2 \in \mathcal{S}_{\mathrm{egress}}$ yields a valid cross-function path $v_2 \xrightarrow{c_3} v_3 \xrightarrow{c_1} op_2$, with $c_3$ and $c_1$ preserved as edge annotations encoding control-flow context.

The analysis results in a set of fund-flow paths, each represented as a variable-level dependency chain annotated with its associated conditions. 
These conditions, expressed in NL, are recorded for subsequent reasoning but not enforced during traversal, balancing semantic interpretability and scalability for downstream large language model-based analysis.
\section{Uncertainty Quantification via Probing and Fusion}
\label{sec:uncertainty}

To better quantify uncertainty, we probe the model from a general-purpose view capturing global semantics and an attack-specific view focusing on adversarial behaviors. In addition, we introduce misleading prompts as controlled perturbations to examine confidence stability.

In borderline scenarios, a single direct query often misclassifies benign contracts as adversarial (see \baselinename's result in \S\ref{sec:stage1_ablation}), as LLMs tend to be overconfident~\cite{xiong2024can,zhao2024explainability}. In black-box senario, internal confidence signals are inaccessible, making it impossible to directly calibrate the model’s confidence.
When the LLM outputs verbalized confidence scores, the distribution reflects its certainty. A uniform distribution (each of $n$ labels assigned $1/n$) indicates high uncertainty, while a highly skewed one (e.g., 95\% versus 5\%) implies strong confidence. This aligns with the concept of entropy in information theory, which quantifies uncertainty that lower entropy indicates higher reliability.


By integrating multi-dimensional information about the contract and entropy-based uncertainty quantification, we fuse predictions into more stable and robust results. All queries are executed in parallel, and the fusion step maintains an acceptable overall cost of roughly \$0.003 per contract.

\subsection{Binary Label Expansion}

In binary adversarial detection, LLMs often produce overconfident predictions, particularly for borderline cases that exhibit partial suspicious behaviors. This overconfidence can obscure the model’s uncertainty and lead to unreliable decisions. To capture more fine-grained confidence information, we expand the binary classification task into a four-level label spectrum:
\begin{align*}
\setlength{\abovedisplayskip}{3pt}
\setlength{\belowdisplayskip}{3pt} 
\mathcal{L} = \{ \texttt{adversarial}, \texttt{suspicion}, \texttt{uncertain}, \texttt{benign} \}.
\end{align*}
Directly using verbalized confidence scores has been shown to be unreliable. In contrast, comparative judgments based on relative rankings often yield more consistent and trustworthy results~\cite{xiong2024can}. Therefore, we prompt the LLM to rank the four label options by likelihood and assign confidence scores that sum to 100\%, ensuring both interpretability and comparability. An illustrative example of this prompt is shown in \figurename~\ref{fig:stage2_label_bounding}.
This structured output not only enables precise quantification of suspicion levels, but also provides the necessary data foundation for the subsequent entropy-based multi-view fusion.

\begin{figure}[]
    \centering
    \begin{tcolorbox}[title=Prompt Template of Stage II, fontupper=\scriptsize]

        \blue{\{Different contract information: general / specific\}}

        You are analyzing a smart contract deployed on the blockchain.

        Provide your 4 best guesses and the probability that each is correct (0\% to 100\%) for the following question. 
        Give your step-by-step reasoning in a few words first, and then give the final answer using the following format:

        \vspace{-2pt}
        \begin{verbatim}
G1: <ONLY the option letter of first most likely
 guess; not a complete sentence, just the guess!>
P1: <ONLY the probability that G1 is correct, without
 any extra commentary whatsoever; just the
 probability!>
...
G4: <ONLY the ..., just the guess!>
P4: <ONLY the ...; just the probability!>
        \end{verbatim}
        \vspace{-8pt}

        The four options represent different levels of suspicion about whether the contract is an adversarial contract:

        - (A) adversarial: clearly malicious, with strong evidence of harmful or exploitative behavior.

        - (B) suspicion: likely malicious, showing noticeable but not definitive attack patterns.

        - (C) uncertain: appears non-malicious but lacks enough evidence to be fully confident.

        - (D) benign: clearly non-malicious, safe with no signs of harmful behavior.

        Options:

        (A) adversarial (B) suspicion (C) uncertain (D) benign

        \blue{(Optional)} Hint: I think the answer should be \{(A) adversarial/(D) benign\}.

    \end{tcolorbox}
    \caption{Prompt Template of Stage II. The contract information of general-purpose analysis is provided in \figurename~\ref{fig:stage2_prompt_contract}, and the attack-specific information is provided in \figurename~\ref{fig:stage2_prompt_function}.}
    \label{fig:stage2_label_bounding}
\end{figure}

\subsection{Confidence Probing}

In Stage I, we obtained both general-purpose and attack-specific information regarding the contract. Due to their abstract nature, general-purpose information often fail to capture concrete behavioral patterns, thereby biasing the model toward benign classifications. In contrast, attack-specific information provides detailed contextual cues more commonly associated with malicious behavior. However, such information may not always clearly indicate attacking behavior due to its specificity and may bias the model towards malicious judgment in some cases. Hence, incorporating the both perspectives is essential for a more comprehensive behavioral analysis.

Furthermore, prior research has shown that misleading or weakly suggestive prompts can influence the model's expressed confidence, even when the final prediction remains unchanged. For example, if the LLM is highly confident that a contract is adversarial, introducing a misleading cue implying benign behavior has minimal effect. Conversely, when the model is uncertain, a misleading cue suggesting adversarial intent can increase its confidence in classifying the contract as adversarial. Building on this insight, we propose the use of misleading prompts to probe the LLM's confidence expression.


Specifically, we design six types of prompts to evaluate the model's confidence across diverse perspectives, defined as:
\begin{align*}
\setlength{\abovedisplayskip}{3pt}
\setlength{\belowdisplayskip}{3pt} 
\mathcal{P} = \{ &\texttt{g\_normal}, \texttt{s\_normal}, \texttt{g\_mislead\_adv}, \\
&\!\!\!\!\!\!\!\!\!\!\!\!\texttt{g\_mislead\_be}, \texttt{s\_mislead\_adv}, \texttt{s\_mislead\_be} \},
\end{align*}
where \(\texttt{g\_normal}\) and \(\texttt{s\_normal}\) correspond to general-purpose and attack-specific information, respectively; \(\texttt{g\_mislead\_adv}\) and \(\texttt{g\_mislead\_be}\) denote adversarial and benign misleading prompts based on general-purpose information; while \(\texttt{s\_mislead\_adv}\) and \(\texttt{s\_mislead\_be}\) apply similar misleading cues to attack-specific information. Each misleading prompt contains a soft hint such as: ``\textit{Hint: I think the answer should be (A) adversarial / (D) benign}'', as illustrated in \figurename~\ref{fig:stage2_label_bounding}.

\subsection{Entropy-Based Fusion and Decision}

Confidence entropy provides a principled measure of uncertainty in a prediction. Higher entropy indicates greater ambiguity in the model's output, while lower entropy suggests more stable and reliable predictions, consistent with Shannon's information theory~\cite{shannon1948mathematical}. 
Based on this principle, we adopt an entropy-based fusion strategy to aggregate multi-view outputs obtained from six different prompts. Specifically, each view produces a ranked confidence distribution over the label options, and we compute the entropy of each distribution. During fusion, predictions from views with lower entropy are assigned higher weights, reflecting their higher reliability. This approach effectively leverages the consistency across multiple prompts to improve classification robustness without incurring significant computational overhead.

Based on this principle, we adopt an entropy-based fusion strategy to aggregate multi-view outputs from six prompts. Each prompt instructs the LLM to assign confidence scores to four label options, forming a normalized confidence distribution. The entropy of this distribution measures view-level uncertainty: a lower entropy indicates more decisive predictions. During fusion, views with lower entropy receive higher weights, thereby emphasizing more reliable judgments while maintaining negligible computational overhead.

Each of the 6 prompts generates a confidence distribution over 4 labels, denoted as:
\begin{align}
\setlength{\abovedisplayskip}{3pt}
\setlength{\belowdisplayskip}{3pt} 
\left\{ \big( G_i^{(p)}, P_i^{(p)} \big) \mid p \in \mathcal{P},\; i \in \mathcal{L},\; \sum_{i \in \mathcal{L}} P_i^{(p)} = 100\% \right\},
\end{align}
where \( G_i^{(p)} \) is the label \( i \) under the prompt \( p \), and \( P_i^{(p)} \) is the corresponding confidence score, which sums to 100\% across all labels for each prompt.
To quantify the uncertainty of each prompt's output, we compute the entropy as
\begin{align}
\setlength{\abovedisplayskip}{3pt}
\setlength{\belowdisplayskip}{3pt} 
H^{(p)} &= - \sum_{i \in \mathcal{L}} p_i^{(p)} \log p_i^{(p)},
\end{align}
where \(H^{(p)}\) denotes the entropy score corresponding to the prompt \(p\).
A lower entropy implies a more confident prediction. Hence, we define the weight of prompt \(p\) as:
\begin{align}
\setlength{\abovedisplayskip}{3pt}
\setlength{\belowdisplayskip}{3pt} 
w^{(p)} = \frac{1}{H^{(p)} + \varepsilon}, \quad \varepsilon = 10^{-6}.
\end{align}

Each prompt also produces a ranked list over the 4 labels. We assign discrete scores to the labels based on their ranks, with the highest-ranked label receiving 3 points, followed by 2, 1, and 0. Let \( s_i^{(p)} \) denote the score assigned to label \( i \in \mathcal{L} \) under prompt \( p \in \mathcal{P} \). The final weighted score for each label is computed as:
\begin{align}
\setlength{\abovedisplayskip}{3pt}
\setlength{\belowdisplayskip}{3pt} 
S_i = \sum_{p \in \mathcal{P}} w^{(p)} \cdot s_i^{(p)}.
\end{align}
We then normalize these scores to a probability distribution:
\begin{align}
\setlength{\abovedisplayskip}{3pt}
\setlength{\belowdisplayskip}{3pt} 
\hat{S}_i = \frac{S_i}{\sum_{j \in \mathcal{L}} S_j}.
\end{align}

Finally, to make a binary decision, we merge the four labels into two groups. We compute:
\begin{align}
\setlength{\abovedisplayskip}{3pt}
\setlength{\belowdisplayskip}{3pt} 
\texttt{adv\_score} &= \hat{S}_{\texttt{adversarial}} + \hat{S}_{\texttt{suspicion}}, \\
\texttt{be\_score} &= \hat{S}_{\texttt{benign}} + \hat{S}_{\texttt{uncertain}}.
\end{align}
If \(\texttt{adv\_score} > \texttt{be\_score}\), the contract is classified as \texttt{adversarial}; otherwise, it is classified as \texttt{benign}.

\subsection{Adaptive Detection Trade-offs}
\label{sec:trade_off}


Thanks to our flexible aggregation strategy, we obtain a continuous adversarial score (\texttt{adv\_score}) rather than a binary label, enabling finer perception of the LLM’s confidence. Users can select thresholds according to their needs: a lower threshold favors sensitivity (lower FNR), while a higher one reduces false alarms (lower FPR). Contracts with scores above the threshold are classified as adversarial, and others as benign.
The corresponding trade-offs between sensitivity and specificity under different thresholds are illustrated in \S\ref{sec:eval_threshold}, with further quantitative details provided in the experimental section.

\section{Evaluation}

\subsection{Experimental Setup}

\subsubsection{Dataset}
\label{sec:dataset}

To comprehensively evaluate \name, we constructed two datasets: $\mathcal{D}_{GT}$, a ground-truth-labeled dataset, and $\mathcal{D}_{RW}$, a collection of contracts from the wild.

For the adversarial contracts in $\mathcal{D}_{GT}$, due to the limited and fragmented availability of labeled adversarial contract data, we made our best effort to comprehensively collect relevant information and annotate it through a combination of original source labels and expert refinement. Specifically, we gathered data from six sources~\cite{yang2024uncover,wang2024skyeye,zhang2025following,defihacklabs2025,pcaversaccio_reentrancy_attacks,ren2025lookahead}, resulting in a total of 455 adversarial contracts across multiple chains, including Ethereum, BSC, Optimism and Polygon.
Among these, 200 contracts contained initial type labels that could be traced to reported attack incidents. We further refined and standardized these labels by referencing incident reports (e.g., tweets, blogs, and disclosures) and analyzing the decompiled pseudocode of the corresponding adversarial contracts. The normalized labels were aligned with the OWASP Smart Contract Top 10 (2025)\footnote{\url{https://owasp.org/www-project-smart-contract-top-10/}}
, a widely recognized guideline that categorizes the most critical smart contract vulnerabilities.
We focus on attacks actively initiated by adversarial contracts, excluding cases such as rug pulls and private key compromises. As shown in \tablename~\ref{table:dataset_comparison}, the labeled subset of $\mathcal{D}_{GT}$, consisting of 200 adversarial contracts with normalized vulnerability categories, represents the only publicly available and the largest multi-type labeled dataset of adversarial contracts to date. The complete set of 455 adversarial contracts further serves as the largest and most comprehensive dataset covering diverse attack types.
For the benign subset of $\mathcal{D}_{GT}$, we utilized the benign contracts provided by Lookahead and randomly sampled 20,000 contracts.

As for $\mathcal{D}_{RW}$, we collected contract data from BscScan over a 10-day period from April 18 to April 27 (UTC time), 2025. We extracted transaction information and contract creation records to obtain contract addresses, and subsequently crawled the corresponding runtime bytecode. Contracts that had already self-destructed were excluded from the dataset. As a result, we obtained 13,210 valid contracts. This dataset was only used in \S\ref{sec:real_world}.

\subsubsection{Metrics}


We use the Balanced Accuracy (BAC)\footnote{\url{https://scikit-learn.org/stable/modules/generated/sklearn.metrics.balanced_accuracy_score.html}} to evaluate detection performance.
Balanced accuracy is particularly suitable for binary and multiclass classification problems~\cite{brodersen2010balanced, mosley2013balanced, amin2016comparing} involving imbalanced datasets. It is defined as the average of recall values across all classes, and is particularly suitable for binary and multiclass classification problems involving imbalanced datasets and has been widely adopted in detection tasks on imbalanced datasets~\cite{wu2024semantic}.
To provide a more comprehensive evaluation, we also report the False Positive Rate (FPR), False Negative Rate (FNR), True Positive Rate (TPR), and True Negative Rate (TNR).

\subsubsection{Baseline Selection}


To our knowledge, only two prior works perform adversarial contract detection directly from bytecode without targeting specific attack types:
\begin{itemize}
    \item \textbf{Lookahead~\cite{ren2025lookahead}:} Uses 40 handcrafted features (on-chain transactions and static bytecode features) and a PSCFT module to extract semantic information from bytecode. Both are inputs to ML classifiers.
    \item \textbf{Skyeye~\cite{wang2024skyeye}:} Selects 12 features from Lookahead and applies control flow graph (CFG) segmentation for semantic extraction. These representations are also used with ML models. We reproduced their method from partially released code.
\end{itemize}
For fairness, we followed their setups by oversampling adversarial contracts, using an 80/20 train–test split, and training for 10 epochs (Lookahead) and 20 epochs (Skyeye).
We also add \textbf{\baselinename} in \S\ref{sec:stage1_ablation} to compare our method against directly feeding NL descriptions to an LLM.

\subsubsection{Implementation}
\label{sec:implementation}



Experiments were conducted on an Ubuntu 24.04 server with a 64-core Intel Xeon Gold 6426Y CPU, eight NVIDIA RTX 4090 GPUs (24GB each), and 256GB RAM. 
\name\ uses \texttt{deepseek-v3}~\cite{liu2024deepseek} as the default LLM, and we further evaluate generalizability using \texttt{claude-3-5-haiku-20241022}~\cite{anthropic2023claude} and \texttt{gpt-4o-2024-08-06}~\cite{openai_gpt4o}.

\subsubsection{Research Questions}

\begin{itemize}
    \item \textbf{RQ1:} How does \name compare to existing baseline methods in detecting adversarial contracts under a training-free paradigm? \textbf{(Baseline Comparison)}
    
    \item \textbf{RQ2:} What is the impact of individual components in the Stage I probing module on detection performance? \textbf{(Ablation Study of Stage I)}
    
    \item \textbf{RQ3:} What is the impact of each component in the Stage II fusion and reasoning module? \textbf{(Ablation Study of Stage II)}
    
    \item \textbf{RQ4:} How efficient and adaptable is \name in terms of inference time, token cost, cross-LLM generalization, and threshold flexibility? \textbf{(Scalability and Efficiency)}
    
    \item \textbf{RQ5:} How effective is \name in detecting adversarial contracts in real-world settings, and what is its potential impact in practical deployments? \textbf{(Real-World Impact)}
\end{itemize}

\subsection{RQ1: Baseline Comparison}

To ensure fair comparison and reliable results, we use the same training and testing splits across all experiments. Specifically, we randomly select 20\% of adversarial and benign contracts from $\mathcal{D}_{GT}$ as the testing set, with the remaining 80\% used for training. This process is repeated five times, and the average performance is reported. All baseline methods are trained and evaluated using these five data splits, except for the ``unseen pattern'' experiment in \S\ref{sec:eval_new_type}, which requires special consideration of adversarial contract types. As our method does not require training, we directly evaluate it on the five testing sets.

\subsubsection{Overall Performance}
\label{sec:eval_overall_performance}

\tablename~\ref{tab:overall_performance} summarizes the detection performance of different methods. We observe that \name achieves a state-of-the-art balanced accuracy (BAC) of 0.9374, outperforming Lookahead (0.8235) and Skyeye (0.9049). The false negative rates (FNR) of Lookahead and Skyeye are 0.3516 and 0.1868, respectively, indicating their high miss rates in detecting adversarial contracts. In contrast, \name significantly reduces the FNR to 0.0769, demonstrating its superior capability in identifying attack behaviors.
The ``Training Time'' column in \tablename~\ref{tab:overall_performance} represents the total time required for model training after all necessary data preparation has been completed.
Specifically, Lookahead requires up to 16.58 minutes, while Skyeye takes approximately 1,440 minutes (about 24 hours). In comparison, \name is a training-free approach that requires no training time, making it significantly more efficient.









\begin{table}[t]
\centering
\setlength{\tabcolsep}{2pt}
\caption{Overall performance of different methods. Training time is reported in minutes.}
\label{tab:overall_performance}

\begin{tabularx}{\linewidth}{
    l
    >{\centering\arraybackslash}X
    >{\centering\arraybackslash}X
    >{\centering\arraybackslash}X
    >{\centering\arraybackslash}X
    >{\centering\arraybackslash}X
    >{\centering\arraybackslash}X
}
\toprule
\multirow{2}{*}{\textbf{Method}}
& \multirow{2}{*}{\textbf{FPR~$\downarrow$}}
& \multirow{2}{*}{\textbf{FNR~$\downarrow$}}
& \multirow{2}{*}{\textbf{TPR~$\uparrow$}}
& \multirow{2}{*}{\textbf{TNR~$\uparrow$}}
& \multirow{2}{*}{\textbf{BAC~$\uparrow$}}
& \textbf{Training} \\
& & & & & & \textbf{Time~$\downarrow$} \\
\midrule

\multirow{1}{*}{Lookahead}
& 0.0014 & 0.3516 & 0.6484 & 0.9987 & 0.8235 & 16.58 \\
\multirow{1}{*}{Skyeye}
& 0.0034 & 0.1868 & 0.8132 & 0.9966 & 0.9049 & 1440.06 \\
\multirow{1}{*}{\textbf{\name}}
& 0.0483 & \textbf{0.0769} & \textbf{0.9231} & 0.9517 & \textbf{0.9374} & \textbf{0.00} \\
\bottomrule

\end{tabularx}
\end{table}

\subsubsection{Unseen Pattern Evaluation}
\label{sec:eval_new_type}


In this experiment, we use the subset of 200 adversarial contracts with normalized type labels in $\mathcal{D}_{GT}$, which covers nine types of adversarial contract. In each experiment, one type \(type_i\) is used as the test set.
To ensure fairness, any contracts that are labeled with multiple types including \(type_i\) are excluded from the test set.
If \(type_i\) has over 10 contracts, 10 are randomly selected for testing; otherwise, all are used. The training set is drawn from the remaining types, capped at 100 contracts if available. This maintains a roughly 100:10 training-to-testing ratio. We report per-type results along with micro- and macro-averaged metrics to provide a comprehensive evaluation.

\tablename~\ref{tab:new_type} summarizes the results of baseline methods and \name under the new attack type evaluation.  
Lookahead's FNR rises sharply to 0.6610 (micro) and 0.7043 (macro), reducing BAC from 0.8235 to 0.6692 and 0.6476, respectively.  
Skyeye shows a similar drop, with BAC decreasing from 0.9049 to 0.7447 (micro) and 0.7529 (macro), and FNR rising to over 0.49.  
In contrast, \name achieves the best BAC of 0.9233, with a low FNR of 0.1050, indicating strong generalization to unseen attack types.
For example, when testing on Type 5 (reentrancy), Lookahead detects only 1 of 10 cases, and Skyeye detects 3, while \name performs significantly better.

\begin{table}[]
\centering
\caption{Performance under unseen attack types. ``micro'' and ``macro'' denote different aggregation methods over the nine attack categories.}
\label{tab:new_type}
\begin{tabular}{lccccc}
\toprule
\textbf{Method}  & \textbf{FPR~$\downarrow$} & \textbf{FNR~$\downarrow$} & \textbf{TPR~$\uparrow$} & \textbf{TNR~$\uparrow$} & \textbf{BAC~$\uparrow$} \\
\midrule
Lookahead (micro) & 0.0006 & 0.6610 & 0.3390 & 0.9994 & 0.6692 \\
Lookahead (macro) & 0.0004 & 0.7043 & 0.2957 & 0.9996 & 0.6476 \\
Skyeye (micro)    & 0.0022 & 0.5085 & 0.4915 & 0.9978 & 0.7447 \\
Skyeye (macro)    & 0.0028 & 0.4914 & 0.5086 & 0.9972 & 0.7529 \\
\textbf{\name}     & 0.0484 & \textbf{0.1050} & 0.8950 & 0.9517 & \textbf{0.9233} \\
\bottomrule
\end{tabular}
\end{table}

\subsubsection{Training vs. Training-Free Performance}
\label{sec:eval_train_ratio}

To evaluate the impact of training set size on method performance, we conducted stratified sampling on the pre-partitioned training dataset, keeping the train-test split fixed for fair comparison.

\tablename~\ref{tab:training_ratio} presents results across various sampling ratios. Even at 10\% (1,636 contracts in training set), the training set remains larger than many datasets used by existing ML methods~\cite{wu2024semantic}. As training data decreases from 100\% to 10\%, Lookahead’s BAC drops from 0.8235 to 0.7604, Skyeye’s from 0.9049 to 0.8488, while \name consistently maintains a BAC of 0.9374, demonstrating robustness as a training-free method.


Notably, Lookahead and Skyeye require 40 and 12 hand-crafted features respectively, and the process of obtaining these features incurs \textbf{substantial extra cost} that is not included in the training time reported in \tablename~\ref{tab:training_ratio}. In contrast, \name only requires the contract bytecode as input, eliminating the need for any extra data preparation. 

\begin{table}[t]
\centering
\setlength{\tabcolsep}{2pt}
\caption{Training vs. training-free performance.}
\label{tab:training_ratio}
\begin{tabularx}{\linewidth}{
    l
    >{\centering\arraybackslash}X
    >{\centering\arraybackslash}X
    >{\centering\arraybackslash}X
    >{\centering\arraybackslash}X
    >{\centering\arraybackslash}X
    >{\centering\arraybackslash}X
}
\toprule
\multirow{2}{*}{\textbf{Method}} 
& \multicolumn{1}{c}{\textbf{Sample}} 
& \multirow{2}{*}{\textbf{FPR~$\downarrow$}} 
& \multirow{2}{*}{\textbf{FNR~$\downarrow$}} 
& \multirow{2}{*}{\textbf{TPR~$\uparrow$}} 
& \multirow{2}{*}{\textbf{TNR~$\uparrow$}} 
& \multirow{2}{*}{\textbf{BAC~$\uparrow$}}
\\
& \multicolumn{1}{c}{\textbf{Ratio}} 
& & & & & \\
\midrule

\multirow{6}{*}{Lookahead} 
& 100\% & 0.0014 & 0.3516 & 0.6484 & 0.9987 & 0.8235 \\
& 90\%  & 0.0015 & 0.3626 & 0.6374 & 0.9986 & 0.8180 \\
& 70\%  & 0.0008 & 0.3736 & 0.6264 & 0.9992 & 0.8128 \\
& 50\%  & 0.0021 & 0.3758 & 0.6242 & 0.9980 & 0.8111 \\
& 30\%  & 0.0021 & 0.4352 & 0.5648 & 0.9980 & 0.7814 \\
& 10\%  & 0.0023 & 0.4769 & 0.5231 & 0.9978 & 0.7604 \\
\midrule

\multirow{6}{*}{Skyeye} 
& 100\% & 0.0034 & 0.1868 & 0.8132 & 0.9966 & 0.9049 \\
& 90\%  & 0.0043 & 0.1868 & 0.8132 & 0.9957 & 0.9044 \\
& 70\%  & 0.0034 & 0.2000 & 0.8000 & 0.9967 & 0.8983 \\
& 50\%  & 0.0035 & 0.2022 & 0.7978 & 0.9965 & 0.8972 \\
& 30\%  & 0.0039 & 0.2615 & 0.7385 & 0.9961 & 0.8673 \\
& 10\%  & 0.0057 & 0.2967 & 0.7033 & 0.9943 & 0.8488 \\
\midrule

\textbf{\name} 
& \textbf{0\%} & 0.0483 & \textbf{0.0769} & 0.9231 & 0.9517 & \textbf{0.9374} \\
\bottomrule

\end{tabularx}
\end{table}

\subsubsection{Obfuscation Evaluation}
\label{sec:eval_obfuscation}

Lookahead and Skyeye depend on on-chain transaction features like \emph{Nonce}, \emph{Value}, \emph{InputDataLength}, \emph{GasUsed}, and \emph{Verified}, which are easily manipulated through simple obfuscation.  
Ma et al.~\cite{ma2023abusing} revealed that verified Ethereum contracts’ bytecode and source code may mismatch, exposing verification-breaking vulnerabilities with real exploitable cases.  
Inspired by this, we simulate an attacker submitting benign-appearing source code that doesn’t match deployed bytecode, setting the Verified feature to ``True'' for adversarial contracts to evade detection.  

\tablename~\ref{tab:obf} shows that Lookahead and Skyeye’s FNRs increase (from 0.3516 to 0.5956 and from 0.1868 to 0.2132), and BACs drop to 0.7017 and 0.8917. In contrast, \name, which doesn’t rely on on-chain features, remains unaffected.  
Other features like \emph{Nonce} can also be obfuscated by sending benign transactions before deploying the real adversarial contract, disguising deployment patterns.

\begin{table}[htbp]
\centering
\caption{Performance under obfuscation scenarios.}
\label{tab:obf}
\begin{tabular}{lccccc}
\toprule
\textbf{Method} & \textbf{FPR~$\downarrow$} & \textbf{FNR~$\downarrow$} & \textbf{TPR~$\uparrow$} & \textbf{TNR~$\uparrow$} & \textbf{BAC~$\uparrow$} \\
\midrule
Lookahead & 0.0011  & 0.5956  & 0.4044  & 0.9990  & 0.7017  \\
Skyeye    & 0.0034  & 0.2132  & 0.7868  & 0.9966  & 0.8917  \\
\textbf{\name}      & 0.0483  & \textbf{0.0769}  & 0.9231  & 0.9517  & \textbf{0.9374}  \\
\bottomrule
\end{tabular}
\end{table}

\begin{tcolorbox}[size=title, opacityfill=0.1, breakable]
\textbf{Summary to RQ1}:
\name achieves a SOTA BAC of 0.9374 and low FNR of 0.0769, significantly better than baselines. \name remains robust under challenging scenarios where baseline performance degrades rapidly.
\end{tcolorbox}

\subsection{RQ2: Impact of Behavioral Semantics}
\label{sec:stage1_ablation}



To evaluate the effectiveness of behavioral semantics, we introduce an additional baseline:
\begin{itemize}
    \item \textbf{\baselinename:} NL descriptions are fed directly into the LLM with a task-specific prompt (see \figurename~\ref{fig:disco_nl_prompt}).
\end{itemize}
We also conduct an ablation study by removing the general-purpose view (\S\ref{sec:attack_intent}, \name~w/o G) and the attack-specific view (\S\ref{sec:attack_action}, \name~w/o S) to assess their individual contributions.

As shown in \tablename~\ref{tab:stage1_ablation}, removing the general-purpose view increases FPR to 0.2341, while removing the attack-specific view raises FNR to 0.1473. The corresponding BAC drops to 0.8532 and 0.9010, respectively. \baselinename performs worst, with an FPR of 0.5712 and BAC of 0.7067.  
These results highlight the importance of both views for accurate detection.

\begin{table}[htbp]
\centering
\caption{Performance comparison of different input information for Stage I.}
\label{tab:stage1_ablation}
\begin{tabular}{lccccc}
\toprule
\textbf{Method}  & \textbf{FPR~$\downarrow$} & \textbf{FNR~$\downarrow$} & \textbf{TPR~$\uparrow$} & \textbf{TNR~$\uparrow$} & \textbf{BAC~$\uparrow$}  \\
\midrule
\name~w/o G   & 0.2341  & 0.0595  & 0.9405  & 0.7659  & 0.8532  \\
\name~w/o S   & 0.0508  & 0.1473  & 0.8527  & 0.9493  & 0.9010  \\
\baselinename & 0.5712  & 0.0154  & 0.9846  & 0.4289  & 0.7067 \\
\textbf{\name}     & 0.0483  & 0.0769  & 0.9231  & 0.9517  & \textbf{0.9374}  \\
\bottomrule
\end{tabular}
\end{table}



\subsection{RQ3: Impact of Confidence Probing and Fusion Strategy}
\label{sec:stage2_ablation}

In this section, we evaluate the effectiveness of the confidence probing and fusion strategies in Stage~II. We compare our method with some alternatives: (1) \texttt{g\_normal} and (2) \texttt{s\_normal}, which use a single prompt respectively; (3) \texttt{no\_misleading}, which fuses the two non-misleading prompts in \name; and (4) \texttt{borda count}, which ranks and aggregates confidence scores using the Borda count method~\cite{wiki_borda_count}.  

\name achieves the highest BAC and the lowest FPR among all methods.
Results are summarized in \tablename~\ref{tab:fusion_comparison}. Methods (1)-(3) show the necessity of using all six prompts, while (4) demonstrate the effectiveness of our fusion strategy. This confirms that our approach better integrates multi-perspective information and fuses predictions with varying confidence levels to enhance overall detection performance.


\begin{table}[htbp]
\centering
\caption{Performance under different probing and fusion strategies.}
\begin{tabular}{lccccc}
\toprule
\textbf{Method}  & \textbf{FPR~$\downarrow$} & \textbf{FNR~$\downarrow$} & \textbf{TPR~$\uparrow$} & \textbf{TNR~$\uparrow$} & \textbf{BAC~$\uparrow$}  \\
\midrule
g\_normal               & 0.0643  & 0.1275  & 0.8725  & 0.9358  & 0.9041  \\
s\_normal               & 0.3002  & 0.0573  & 0.9427  & 0.6998  & 0.8213  \\
no\_misleading          & 0.0912  & 0.0703  & 0.9297  & 0.9089  & 0.9193  \\
borda\_count            & 0.1290  & 0.0637  & 0.9363  & 0.8710  & 0.9036  \\
\textbf{\name}      & \textbf{0.0483}  & 0.0769  & 0.9231  & \textbf{0.9517}  & \textbf{0.9374}  \\
\bottomrule
\end{tabular}
\label{tab:fusion_comparison}
\end{table}



\subsection{RQ4: Scalability and Efficiency}

\subsubsection{Optional Threshold Selection}
\label{sec:eval_threshold}

As discussed in \S\ref{sec:trade_off}, users may have different preferences regarding the balance between FPR and FNR depending on specific application scenarios. 
To accommodate this, we provide an optional threshold selection mechanism, allowing users to adjust the adversarial score threshold based on their desired trade-off.
Generally, a higher threshold yields lower FPR but higher FNR, while a lower threshold results in the opposite. 
As shown in \figurename~\ref{fig:metrics_across_threshold}, setting the threshold to 0.18 results in a very low FNR of 0.0088, whereas increasing the threshold to 0.7 reduces the FPR to 0.0003. 
This illustrates the inherent trade-off between sensitivity and specificity in threshold-based decisions.


\subsubsection{Evaluation under Different LLM Backbones}
\label{sec:diff_llm}

To evaluate the generalizability of our approach, we conduct experiments using different LLM backbones, as shown in \tablename~\ref{tab:llm_compare}.
\name achieves consistently strong performance across models, with peak BAC scores of 0.9064 and 0.8461 on \texttt{Claude 3.5} and \texttt{GPT-4o}, and correspondingly low FNRs of 0.1429 and 0.1231 with an optimized threshold, outperforming the baselines by up to 59.36\% and 64.99\% in FNR.
Notably, the best performance is observed on \texttt{DeepSeek-V3}, with a BAC of 0.9374 and an FNR of 0.0769.
These results demonstrate that \name maintains high detection accuracy across a range of LLMs, indicating good generalizability.
In particular, \texttt{Claude 3.5} surpasses all baselines in both BAC and FNR, and \texttt{GPT-4o} delivers comparable performance.

\begin{table}[h]
\centering
\caption{Performance under different LLMs.}
\label{tab:llm_compare}
\begin{tabular}{lccccc}
\toprule
\textbf{Model} & \textbf{FPR~$\downarrow$} & \textbf{FNR~$\downarrow$} & \textbf{TPR~$\uparrow$} & \textbf{TNR~$\uparrow$} & \textbf{BAC~$\uparrow$} \\
\midrule
\textbf{DeepSeek-V3} & 0.0483  & 0.0769  & 0.9231  & 0.9517  & \textbf{0.9374}  \\
Claude 3.5  & 0.0444  & 0.1429  & 0.8571  & 0.9557  & 0.9064  \\
GPT-4o      & 0.1848  & 0.1231  & 0.8769  & 0.8153  & 0.8461  \\

\bottomrule
\end{tabular}
\end{table}


\subsubsection{Efficiency Analysis}


We compare average token and time costs across models to assess efficiency. For \texttt{Deepseek-V3}, the average token usage is 17,085.48, with a per-contract cost of \$0.0030, including \$0.0018 from Stage I and \$0.0012 from Stage II. In comparison, \texttt{Claude-3.5} costs \$0.0021, while \texttt{GPT-4o} costs \$0.0664 per contract.
Since our method operates solely on bytecode, it supports pre-deployment detection and reduces overall overhead.




\subsection{RQ5: Real-World Impact}
\label{sec:real_world}


To evaluate the real-world effectiveness of \name, we collected five publicly reported adversarial contracts (\tablename~\ref{table:real_world}) from reliable sources, all of which were correctly identified based on bytecode, achieving a 100\% TPR. 
Additionally, by randomly sampling contracts from $\mathcal{D}_{RW}$ and validating them with expert assistance, we uncovered 29 previously undocumented adversarial contracts showing clear malicious intent or stealthy behavior.

\subsubsection{Case Study}
\label{sec:case_study}

As part of our case study, we analyze an adversarial contract \href{https://bscscan.com/address/0x75f2002937507b826b727170728595fd45151d0f}{(see BscScan link)} flagged by \name and independently confirmed by TenArmorAlert~\cite{TenArmorAlert2024}.
The attack resulted in a loss of approximately \$26K.
As the source code is unavailable, we integrate two state-of-the-art tools, Disco~\cite{su2025disco} for source code recovery and Elipmoc~\cite{grech2022elipmoc} for pseudocode lifting, together with manual validation to ensure that the recovered logic is semantically consistent with the on-chain runtime bytecode, as shown in \figurename~\ref{fig:case_study_code}.

As analyzed in \textbf{Obs2}, this contract executes its attack in typical three stages: preparation (injecting funds via \texttt{exploit()}), exploitation (arbitrage via \texttt{DPPFlashLoanCall()}), and exit (extracting illicit gains in \texttt{DPPFlashLoanCall()}).
\name extracted key fund-flow paths (\figurename~\ref{fig:motivation_case_taint_paths}).
\tablename~\ref{tab:case_study_prompt_result} shows prediction scores from 6 prompts. After entropy-based fusion, the adversarial, suspicion, uncertain, and benign scores are 0.288, 0.327, 0.221, and 0.163, respectively. The combined adversarial score (adversarial + suspicion) is 0.615, leading to a final prediction of adversarial.

\subsubsection{Real World Findings}

Notably, \name identified instances of price manipulation and arbitrage facilitated by flash loans, as illustrated in \figurename~\ref{fig:case_study_code}. These cases represent a rapidly emerging and increasingly prevalent threat in recent years, distinctly different from conventional flash loan attacks that exploit protocol vulnerabilities, as well as from benign arbitrage strategies that leverage price inefficiencies without adversarial manipulation. The detected behaviors involve adversaries borrowing large amounts of assets via flash loans to actively distort price dynamics across protocols and extract profits through carefully orchestrated arbitrage. These cases conform to the three-stage adversarial pattern abstracted in our model, highlighting \name's capability to capture emerging yet previously unseen attack patterns.

\section{Discussion}




\textbf{Limitations.}
Our method detects adversarial behavior exhibited directly by contracts. It may miss proxy contracts that delegate malicious logic elsewhere, as these proxies show no adversarial actions themselves. For example, some false negatives occur when the actual exploit lies in a downstream callee, not the proxy.

\textbf{Future Work.}
Currently, we analyze adversarial contracts independently, focusing on internal logic and fund flows. However, many attacks involve subtle interactions with victim contracts. Future work could integrate victim contracts into the analysis, enabling joint reasoning over attacker and target behaviors.
Since most adversarial contracts hardcode victim addresses or store them in fixed slots~\cite{yang2024uncover}, cross-contract analysis is both feasible and promising. This would improve detection of coordinated attacks (e.g., reentrancy, storage collisions), track fund flows across contracts, and enhance accuracy in complex scenarios.


\section{Related Work}

\subsection{Vulnerability Detection in Smart Contracts}
\label{sec:vuln_detection}

Vulnerability detection is a long-standing and critical task in the blockchain ecosystem.  
Traditional methods can be broadly categorized into four types:  
static analysis-based~\cite{feist2019slither},  
symbolic execution-based~\cite{wang2020artemis},  
fuzzing-based~\cite{torres2021confuzzius}, and  
machine learning-based approaches~\cite{nguyen2022mando}.
LLMs have recently been explored for vulnerability detection~\cite{lin2025large},  
with hybrid methods combining them with static analysis~\cite{sun2024gptscan}, multi-agent reasoning~\cite{ma2024combining}, or fuzzing~\cite{shou2024llm4fuzz}.

Despite recent progress, existing approaches face three key limitations.  
First, most detected vulnerabilities are not practically exploitable: only 2.68\%~\cite{yang2024uncover} of flagged contracts are actually exploited.  
Second, high false positives remain common~\cite{song2025silence}, such as misclassifying safe anti-reentrancy contracts as vulnerable. 
Third, there remains an inherent asymmetry: defenders must address all threats, while attackers only need to exploit one.
These issues highlight the need for proactive, adversary-aware detection focused on real-world exploits rather than theoretical flaws.

\subsection{Adversarial Detection in Transactions}



Some works aim to detect attacks in real time by monitoring unconfirmed transactions in the public mempool.  
They typically follow two approaches:  
pattern-based~\cite{wang2021towards,wu2023defiranger,xie2024defort}, using manually defined heuristics; and  
learning-based~\cite{babel2023lanturn,li2023demystifying,qian2023demystifying}, leveraging ML models with graph-based features.

While effective in public settings, these methods fail to detect transactions submitted via private mempools, which bypass public monitoring.  
Moreover, their reliance on fixed rules or models limits adaptability to novel or evolving attacks.

\subsection{Adversarial Detection in Smart Contracts}

Beyond vulnerability identification, recent research has focused on detecting adversarial smart contracts deliberately designed to exploit vulnerabilities or manipulate protocol logic. For example, Yang et al.~\cite{yang2024uncover} propose a method to identify contracts capable of launching exploitable reentrancy attacks.
Smartcat~\cite{zhang2025following} proposes an efficient static analyzer that detects price manipulation adversarial contracts via cross-function call graphs and token flow graphs.
However, these approaches target only a specific type of adversarial contracts.

In contrast to pattern-specific detection, two works aim to develop ML-based systems that generalize across various types of adversarial contracts. Lookahead~\cite{ren2025lookahead} lifts bytecode into TAC using Elipmoc~\cite{grech2022elipmoc}, and combines Pruned Semantic-Control Flow Tokenization with handcrafted features using a dual-branch architecture for final classification. Skyeye~\cite{wang2024skyeye} extracts static features and performs control flow graph segmentation to build a unified representation for training a binary classifier.
However, ML-based approaches heavily rely on labeled training data, limiting their ability to detect novel or unseen attack variants. They are trained on low-level opcode features, lacking the capacity to capture high-level semantic intent. Furthermore, their poor interpretability hinders security analysts from understanding model decisions, which is critical in high-stakes security contexts.

\section{Conclusion}

In this work, we presented \name, a novel and training-free framework for detecting adversarial smart contracts directly from EVM bytecode. By lifting low-level bytecode into semi-structured natural language description, \name enhances semantic understanding and enables multi-view analysis of contract behavior.
Additionally, \name incorporates fund-flow reachability analysis to capture the distinct stages of attacks, thereby strengthening semantic precision.
To further improve robustness, we introduced a fine-grained uncertainty quantification mechanism that mitigates hallucinations and enhances detection reliability. Our method demonstrates strong generalization to unseen attack patterns, resilience in low-data settings, and state-of-the-art performance across multiple metrics. These results highlight the potential of LLM-based semantic reasoning for proactive and effective smart contract security. We believe \name offers a practical step toward securing blockchain ecosystems against emerging adversarial threats.

\newpage

\bibliographystyle{IEEEtran}
\bibliography{IEEEabrv,main}

\appendices

\section{Methodology Supplement}
\subsection{Supplementary Prompts in Stage~I and~II}
\label{sec:stage2_prompt_appendix}

\begin{figure}[t!]
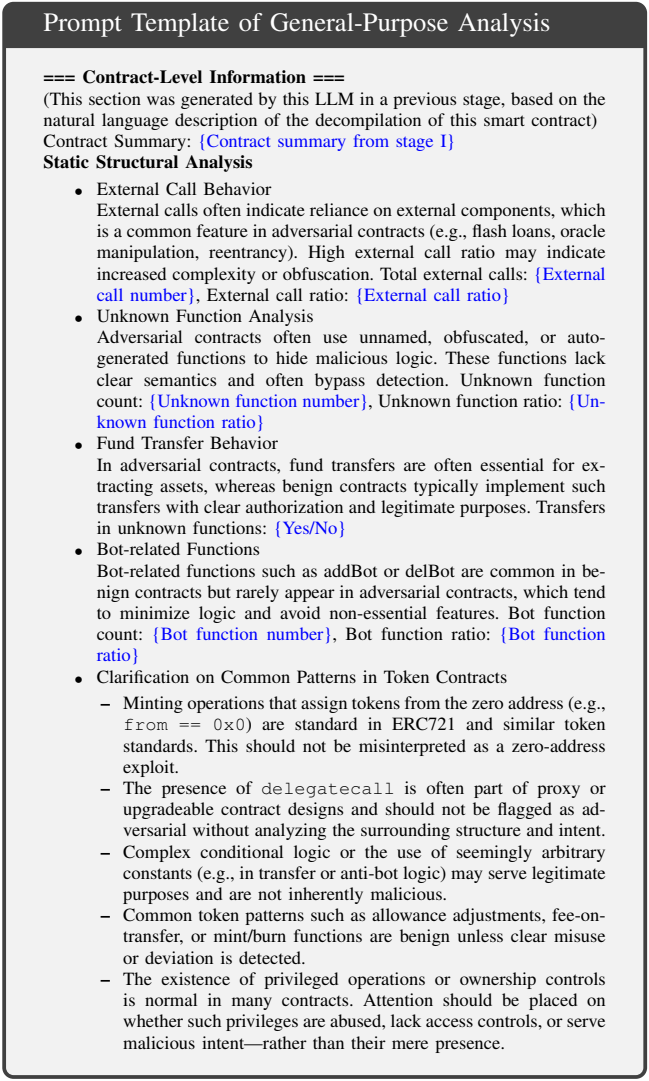

    \begin{tcolorbox}[title=Prompt Template of General-Purpose Analysis, fontupper=\scriptsize]

        \textbf{=== Contract-Level Information ===}\\
        (This section was generated by this LLM in a previous stage, based on the natural language description of the decompilation of this smart contract)

        Contract Summary: \blue{\{Contract summary from stage I\}}

        \textbf{Static Structural Analysis}

        \begin{itemize}
            \item External Call Behavior
            
            External calls often indicate reliance on external components, which is a common feature in adversarial contracts (e.g., flash loans, oracle manipulation, reentrancy).
            High external call ratio may indicate increased complexity or obfuscation.
            Total external calls: \blue{\{External call number\}},             
            External call ratio: \blue{\{External call ratio\}}
            
            \item Unknown Function Analysis
            
            Adversarial contracts often use unnamed, obfuscated, or auto-generated functions to hide malicious logic.
            These functions lack clear semantics and often bypass detection.
            Unknown function count: \blue{\{Unknown function number\}}, 
            Unknown function ratio: \blue{\{Unknown function ratio\}}

            \item Fund Transfer Behavior
            
            In adversarial contracts, fund transfers are often essential for extracting assets, whereas benign contracts typically implement such transfers with clear authorization and legitimate purposes.
            Transfers in unknown functions: \blue{\{Yes/No\}}

            \item Bot-related Functions
            
            Bot-related functions such as addBot or delBot are common in benign contracts but rarely appear in adversarial contracts, which tend to minimize logic and avoid non-essential features.
            Bot function count: \blue{\{Bot function number\}}, 
            Bot function ratio: \blue{\{Bot function ratio\}}

            \item Clarification on Common Patterns in Token Contracts
            \begin{itemize}
                \item Minting operations that assign tokens from the zero address (e.g., \texttt{from == 0x0}) are standard in ERC721 and similar token standards. This should not be misinterpreted as a zero-address exploit.
                \item The presence of \texttt{delegatecall} is often part of proxy or upgradeable contract designs and should not be flagged as adversarial without analyzing the surrounding structure and intent.
                \item Complex conditional logic or the use of seemingly arbitrary constants (e.g., in transfer or anti-bot logic) may serve legitimate purposes and are not inherently malicious.
                \item Common token patterns such as allowance adjustments, fee-on-transfer, or mint/burn functions are benign unless clear misuse or deviation is detected.
                \item The existence of privileged operations or ownership controls is normal in many contracts. Attention should be placed on whether such privileges are abused, lack access controls, or serve malicious intent—rather than their mere presence.
            \end{itemize}
        \end{itemize}

    \end{tcolorbox}
    \caption{The prompt for general-purpose analysis in Stage II.}
    \label{fig:stage2_prompt_contract}
\end{figure}
\begin{figure}[t!]
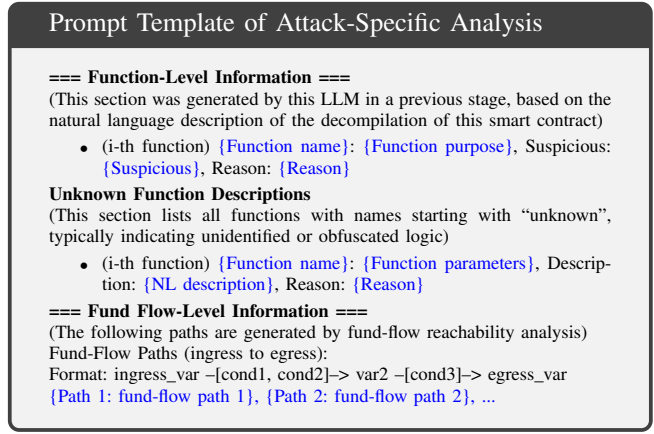

    \begin{tcolorbox}[title=Prompt Template of Attack-Specific Analysis, fontupper=\scriptsize]

        \textbf{=== Function-Level Information ===}\\
        (This section was generated by this LLM in a previous stage, based on the natural language description of the decompilation of this smart contract)

        \begin{itemize}
        \item (i-th function) {\blue{\{Function name\}}}: {\blue{\{Function purpose\}}}, Suspicious: {\blue{\{Suspicious\}}}, Reason: {\blue{\{Reason\}}}
        \end{itemize}
        
        \textbf{Unknown Function Descriptions}\\
        (This section lists all functions with names starting with ``unknown'', typically indicating unidentified or obfuscated logic)

        \begin{itemize}
        \item (i-th function) {\blue{\{Function name\}}}: {\blue{\{Function parameters\}}}, Description: {\blue{\{NL description\}}}, Reason: {\blue{\{Reason\}}}
        \end{itemize}
        
        \textbf{=== Fund Flow-Level Information ===}\\
        (The following paths are generated by fund-flow reachability analysis)

        Fund-Flow Paths (ingress to egress):

        Format: ingress\_var --[cond1, cond2]--> var2 --[cond3]--> egress\_var

        \blue{\{Path 1: fund-flow path 1\},}
        \blue{\{Path 2: fund-flow path 2\},}
        \blue{ ...}

    \end{tcolorbox}
    \caption{The prompt for attack-specific analysis in Stage II.}
    \label{fig:stage2_prompt_function}
\end{figure}

\figurename~\ref{fig:stage1_prompt} shows the prompts for information extraction in Stage~I.
\figurename\ref{fig:stage2_prompt_contract} and \figurename~\ref{fig:stage2_prompt_function} show the general-purpose and attack-specific prompts used in Stage~II.

\begin{figure}[h]
    \begin{tcolorbox}[title=Prompt Template of Stage~I, fontupper=\scriptsize]
        
        \textbf{System:}
        You are an excellent smart contract attack detector. You are given natural language descriptions of smart contracts and are skilled at identifying malicious or suspicious intent. Your task is to generate concise and structured reports highlighting attacker-like behaviors.

        \tcbline

        \textbf{\#\#\# Task: function\_summary \#\#\#}\\
        Please analyze each described function individually, focusing on attacker-like or abnormal behaviors. For each function, provide:
        \begin{enumerate}
            \item {Function Purpose}: A short summary of the function's behavior (1-2 sentences).
            \item {Suspicious Behavior}: Does the function contain signs of attacker-like or exploitative behavior? (Yes/No)
            \item {Evidence (if any)}: Explain why it might be malicious or abnormal (1-2 sentences).
        \end{enumerate}

        \textbf{Response Format (strictly follow):}\\
        \vspace{-13pt}
        \begin{verbatim}
  [Function: <function_name>]
  Purpose: <summary>
  Suspicious Behavior: <Yes or No>
  Evidence: <brief explanation>
        \end{verbatim}
        \vspace{-13pt}

        \tcbline
        
        \textbf{\#\#\# Task: contract\_summary \#\#\#}\\
        Based on the above analysis, please perform a high-level analysis of the smart contract based on its natural language description. Your response must include: briefly summarize the overall purpose and behavior of the contract (40-80 words).

        \textbf{Response Format (strictly follow):}\\
        \vspace{-13pt}
        \begin{verbatim}
  [Contract Summary]
  <Your summary here>
        \end{verbatim}
        \vspace{-13pt}

        \tcbline

        \textbf{\#\#\# Natural Language Description of Smart Contract \#\#\#}\\
        \blue{\{NL description of the smart contract.\}}

    \end{tcolorbox}
    \caption{The prompt for stage~I}
    \label{fig:stage1_prompt}
\end{figure}

\subsection{Behavior Node Types}

\tablename~\ref{table:behavior_types} presents the behavior types and propagation rules used for fund-flow reachability analysis in \S\ref{sec:dependency_forest}.

\begin{table*}[t]
    \centering
    \caption{
    Behavior types and propagation rules for fund-flow reachability. 
    For each type, we define a source entity set \(\mathcal{X}_{\text{src}}\) and a destination entity \(x_{\text{dst}}\) to identify ingress and egress points. 
    Only fund-relevant behaviors are considered, while neutral behaviors such as returns, logs, or built-in calls are excluded as they do not affect fund-flows.
    }

    \label{table:behavior_types}

    \begin{tabularx}{\textwidth}{l X l l}
        \toprule
        \textbf{Behavior Type} & \textbf{NL Description Template} & $\mathbf{\mathcal{X}}_{\mathrm{src}}$ & $\mathbf{x}_{\mathrm{dst}}$ \\
        \midrule
        assignment &
        it updates the state variable $x$ to $y$ &
        $\{ y \}$ (if not constant) &
        $x$ \\

        external call &
        it triggers the external call to contract.function($x_1, x_2, \dots, x_n$) &
        $\{ x_1, x_2, \dots, x_n \}$ (if not constant) &
        $\mathrm{contract.function}$ \\

        delegate call &
        it delegates a call to contract.function($x_1, x_2, \dots, x_n$) &
        $\{ x_1, x_2, \dots, x_n \}$ or \texttt{calldata} (if present) &
        $\mathrm{delegatecall}$ \\

        contract creation &
        it creates a new smart contract with creation code $c$ (and optional salt $s$), and gets a new address $a$ &
        $\{ s \}$ (optional, if not constant) &
        $a$ \\

        transfer &
        it transfers $v$ wei to $a$ (with gas $g$) &
        $\{ v \}$ (if not constant) &
        $\mathrm{transfer}(v)$ \\

        return &
        it returns $x_1, x_2, \dots, x_n$ &
        $\emptyset$ &
        - \\

        log emission &
        it emits the log event with parameter(s) $x_1, x_2, \dots, x_n$ &
        $\emptyset$ &
        - \\

        built-in call &
        it calls a built-in function $f$ &
        $\emptyset$ &
        - \\

        other &
        (fallback for unmatched patterns) &
        $\emptyset$ &
        - \\
        \bottomrule
    \end{tabularx}

    \smallskip
    \raggedright
    \textbf{Note:} In the case of \textit{contract creation}, the variable $s$ denotes the \textit{salt}, which is an optional parameter influencing the new contract address. If not present, the source variable set is empty.
\end{table*}
\subsection{Reachability Analysis Algorithm}

Algorithm~\ref{alg:fund_flow_reachability} presents the reachability analysis used in fund-flow analysis, as described in \S\ref{sec:reachability}.

\begin{algorithm}[]
\caption{Reachability Analysis}
\label{alg:fund_flow_reachability}
\SetAlgoLined
\LinesNumbered
\KwIn{Variable-level dependency graph $G=(V, E)$ with edges labeled by conditions}
\KwOut{Fund-Flow paths $\mathcal{P}$ from ingress to egress}

\mbox{$\mathcal{S}_{ingress}, \mathcal{S}_{egress} \leftarrow$ initial\_ingress\_egress(} \\
\mbox{\quad initialIngress, initialEgress, $\mathcal{F}$)\;}

$\mathcal{E}_{\mathrm{visited}} \leftarrow \{(v, \emptyset) \mid v \in \mathcal{S}_{\mathrm{ingress}}\}$ \tcp*{Visited variables and associated conditions}
$updated \leftarrow$ \textbf{true}\;

\While{$updated$}{
    $updated \leftarrow$ \textbf{false}\;
    \ForEach{$v \in V$ such that $v \in \mathrm{Dom}(\mathcal{E}_{\mathrm{visited}})$}{
        \ForEach{edge $(v \xrightarrow{conds} u) \in E$}{
            \If{$u \notin \mathrm{Dom}(\mathcal{E}_{\mathrm{visited}})$}{
                $\mathcal{E}_{\mathrm{visited}} \leftarrow \mathcal{E}_{\mathrm{visited}} \cup \{(u, conds)\}$\;
                $updated \leftarrow$ \textbf{true}\;
            }
        }
    }
}

$\mathcal{P} \leftarrow \mathrm{ReverseDFSPrune}(G, \mathcal{E}_{\mathrm{visited}}, \mathcal{S}_{\mathrm{ingress}}, \mathcal{S}_{\mathrm{egress}})$\;

\Return{$\mathcal{P}$}
\end{algorithm}

\section{Experimental Supplement}
\subsection{Dataset Detail}
\label{sec:dataset_table}

\tablename~\ref{table:dataset_comparison} compares our dataset $\mathcal{D}_{GT}$ with existing ones, while \tablename~\ref{table:dataset} presents the detailed type distribution, as described in \S\ref{sec:dataset}.

\begin{table}[ht]
\centering
\caption{Comparison of dataset size and label availability with existing adversarial contract datasets.}
\label{table:dataset_comparison}
\begin{tabular}{lccc}
\toprule
\textbf{Dataset} & \textbf{Num} & \textbf{Type Label} & \textbf{Multi-Type} \\
\midrule
BlockWatchdog\cite{yang2024uncover}   & 18  & \CIRCLE & \Circle \\
SmartCat\cite{zhang2025following}        & 84  & \CIRCLE & \Circle \\
\textbf{\boldmath$\mathcal{D}_{GT}$ (part)}   & \textbf{200} & \textbf{\CIRCLE} & \textbf{\CIRCLE} \\
\midrule
Lookahead\cite{ren2025lookahead}       & 375 & \Circle & \CIRCLE \\
Skyeye\cite{wang2024skyeye}          & 174 & \Circle & \CIRCLE \\
\textbf{\boldmath$\mathcal{D}_{GT}$}  & \textbf{455} & \textbf{\LEFTcircle (200/455)} & \textbf{\CIRCLE} \\
\bottomrule
\end{tabular}
\end{table}

\begin{table}[htbp]
\centering
\caption{Different Adversarial Contract Types in $\mathcal{D}_{GT}$.}
\label{table:dataset}
\begin{tabular}{llcr>{\centering\arraybackslash}m{1.5cm}} 
\toprule
\textbf{Type ID} & \textbf{Attack Type} & \textbf{Num} & \textbf{Sum} \\ 
\midrule
Type1  & Access Control Vulnerabilities       & 13  &  \multirow{10}{*}{200\footnotemark}  \\
Type2  & Price Oracle Manipulation            & 109 &      \\
Type3  & Logic Errors                         & 29  &      \\
Type4  & Lack of Input Validation             & 7   &      \\
Type5  & Reentrancy Attacks                   & 53  &      \\
Type6  & Unchecked External Calls             & 2   &      \\
Type7  & Flash Loan Attacks                   & 21  &      \\
Type8  & Integer Overflow and Underflow      & 3   &      \\
Type9  & Insecure Randomness                  & 5   &      \\
Type10 & Denial of Service (DoS) Attacks     & 0   &      \\
\midrule
No Type &  &  & 255 \\

\midrule
Total &  &  & 455 \\

\bottomrule
\end{tabular}
\end{table}
\footnotetext{Some contracts involve multiple attack types, so the total count may not match the sum of all types. No DoS-related contracts were observed during data collection.}
\subsection{Prompt of \baselinename}


\figurename~\ref{fig:disco_nl_prompt} shows the prompt template of \texttt{\baselinename} in \S\ref{sec:stage1_ablation}, where the NL descriptions are directly fed into the LLM.

\begin{figure}[t!]
    \begin{tcolorbox}[title=Prompt Template of \baselinename, fontupper=\scriptsize]

    \blue{\{NL description of smart contract.\}}

    This text provides a natural language description of a smart contract deployed on the blockchain. Based on this description, determine whether the contract is adversarial or benign. Please respond using the following format:

    \begin{verbatim}
1) [[adversarial/benign]]
2) [[Your reasoning for this judgment]]
    \end{verbatim}
    \vspace{-13pt}

    \end{tcolorbox}
    \caption{Prompt template for \baselinename .}
    \label{fig:disco_nl_prompt}
\end{figure}
\subsection{Threshold Selection}

\begin{figure}[]
    \centering
    \includegraphics[width=\linewidth]{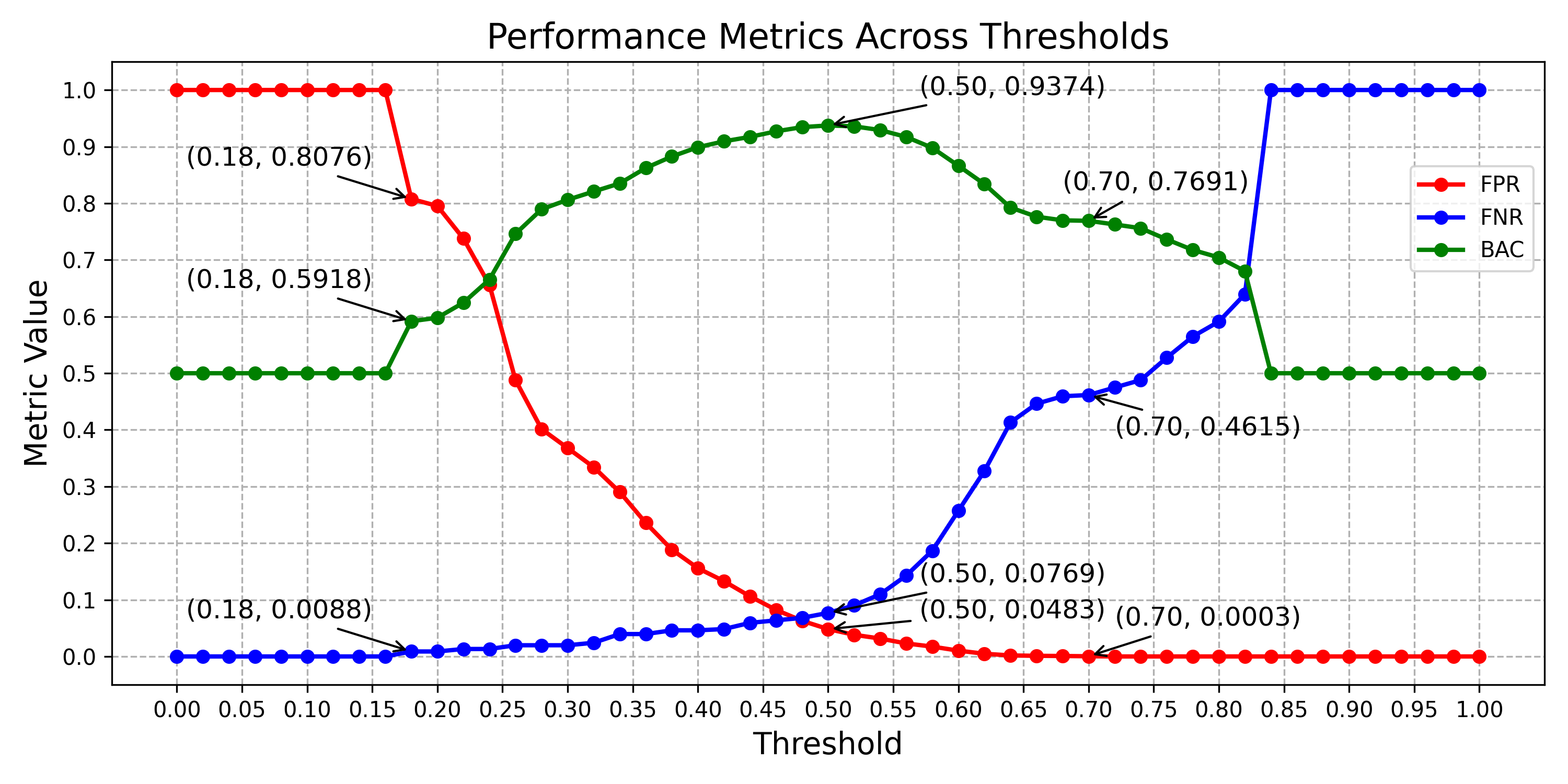}
    \caption{Performance metrics across varying adversarial score thresholds.}
    \label{fig:metrics_across_threshold}
\end{figure}

\figurename~\ref{fig:metrics_across_threshold} presents the experimental results of optional threshold selection in \S\ref{sec:eval_threshold}, showing the variations of FPR, FNR, and BAC with different threshold values.

\subsection{Disclosed Real-World Attack Cases}
\label{sec:appendix_real_world_table}

\tablename~\ref{table:real_world} lists adversarial contracts detected by \name that have been publicly disclosed.

\begin{table}[htbp]
\centering
\setlength{\tabcolsep}{2pt}  
\caption{Real-world attack cases detected by \name and publicly disclosed.}
\label{table:real_world}
\begin{tabular}{cccc}
\toprule
\textbf{Deploy Time} & \textbf{Report} & \textbf{Adversarial Contract Address}\\
\midrule
2025.04.18 & \href{https://x.com/TenArmorAlert/status/1913500336301502542}{link} & 0x7a4d144307d2dfa2885887368e4cd4678db3c27a\\
2025.04.23 & \href{https://x.com/TenArmorAlert/status/1915324379757502727}{link} & 0xdd9a85fd532faadb0c439bbd725e571c4214aedf\\
2025.04.26 & \href{https://x.com/SlowMist_Team/status/1916391258664174045}{link} & 0xf6cee497dfe95a04faa26f3138f9244a4d92f942\\
2025.04.26 & \href{https://x.com/TenArmorAlert/status/1916312483792408688}{link} & 0x75f2002937507b826b727170728595fd45151d0f\\
2025.04.26 & \href{https://bscscan.com/address/0xcfd3cf61619cbec15e9a8bef0e5cd613a565b6b3}{link} & 0xcfd3cf61619cbec15e9a8bef0e5cd613a565b6b3\\
\bottomrule
\end{tabular}
\end{table}

\section{Case Studies}
\subsection{Motivation Example: Understanding \name}
\label{sec:motivation_case}



We provide a detailed analysis of \name, referring to the motivation example in \figurename~\ref{fig:case_study_code} and the earlier case study in \S\ref{sec:case_study}. \figurename~\ref{fig:motivation_case_func} and \figurename~\ref{fig:motivation_case_contract} show the function-centric and contract-centric summaries of this example.

\begin{figure}[htbp]
    \begin{ResultBox}
    \scriptsize\ttfamily
    \begin{itemize}
        \item \textbf{DPPFlashLoanCall}:
            \begin{itemize}
                \item purpose: This function initiates a series of external calls to approve, transfer, buy, and sell tokens, with specific conditions and checks on return data.
                \item suspicious: Yes.
                \item reason: The function requires a specific hash of the transaction origin, which is highly unusual and suggests it may be designed to only allow execution by a specific entity, potentially indicating a backdoor or exploit.
            \end{itemize}
        \item \textbf{receiveFlashLoan}:
            \begin{itemize}
                \item purpose: This function handles the receipt of a flash loan, transferring amounts and checking balances.
                \item suspicious: Yes.
                \item reason: Similar to the previous function, the requirement for a specific hash of the transaction origin suggests restricted access, which could be used to exploit the contract.
            \end{itemize}
        \item \textbf{unknownfffcf3a1}:
            \begin{itemize}
                \item purpose: This function checks a balance and initiates a flash loan with specific parameters.
                \item suspicious: Yes.
                \item reason: The requirement for a specific hash of the transaction origin indicates potential malicious intent, as it restricts function execution to a specific entity.
            \end{itemize}
        
    \end{itemize}

    \end{ResultBox}
    \caption{Function-centric summary of the contract in \figurename~\ref{fig:case_study_code} and Appendix \ref{sec:motivation_case}.}
    \label{fig:motivation_case_func}
\end{figure}

\begin{figure}[htbp]
    \begin{ResultBox}
    \scriptsize\ttfamily

    \begin{itemize}
        \item \textbf{contract summary}: The contract appears to be designed to perform complex financial operations involving flash loans, token transfers, and balance checks. However, it includes specific restrictions on transaction origin, suggesting it may be intended for use by a single entity, potentially for exploitative purposes.
    \end{itemize}

    \end{ResultBox}
    \caption{Contract-centric summary of the contract in \figurename~\ref{fig:case_study_code} and Appendix \ref{sec:motivation_case}.}
    \label{fig:motivation_case_contract}
\end{figure}

\begin{figure}[t]
\begin{ResultBox}
\scriptsize\ttfamily

\textbf{Path 1:}

\hspace*{1.5em}caller --[it is required that (0x268d...4080 == sha3(tx.origin)), it is required that the 2nd external call succeeds, ...]--> stor\_3.transfer

\vspace{0.5em}
\textbf{Path 2:}

\hspace*{1.5em}DPPFlashLoanCall:param1 --[it is required that (0x268d...4080 == sha3(tx.origin)), ..., it is required that the 1st return data of the 3rd external call]--> stor\_3.transfer

\vspace{0.5em}
\textbf{Path 3:}

\hspace*{1.5em}unknownfffcf3a1:param1 --[it is required that (0x268d...4080 == sha3(tx.origin)), ...--> stor\_5.flashLoan

\end{ResultBox}
\caption{Fund-flow analysis of the contract in \figurename~\ref{fig:case_study_code} and Appendix \ref{sec:motivation_case}.}
\label{fig:motivation_case_taint_paths}
\end{figure}

\begin{table}[]
\centering
\begin{tabular}{l c c c c c c c c}
\toprule
\textbf{Prompt} & \textbf{G1} & \textbf{P1} & \textbf{G2} & \textbf{P2} & \textbf{G3} & \textbf{P3} & \textbf{G4} & \textbf{P4} \\
\midrule
prompt\_g\_normal      & B & 60 & A & 25 & C & 10 & D & 5 \\
prompt\_s\_normal      & B & 60 & A & 30 & C & 8 & D & 2 \\
prompt\_g\_mislead\_ad & A & 60 & B & 30 & C & 8 & D & 2 \\
prompt\_g\_mislead\_be & D & 70 & C & 20 & B & 8 & A & 2 \\
prompt\_s\_mislead\_ad & A & 80 & B & 15 & C & 5 & D & 0 \\
prompt\_s\_mislead\_be & D & 60 & C & 30 & B & 10 & A & 0 \\
\bottomrule
\end{tabular}
\caption{Predictions for the contract in \figurename~\ref{fig:case_study_code} and Appendix \ref{sec:motivation_case}.}
\label{tab:case_study_prompt_result}
\end{table}

\subsection{Benign Example: Comparison with \baselinename}
\label{sec:ablation_case}

We conduct a detailed comparison using a benign contract (\href{https://etherscan.io/address/0x5c1bfab1f6bbd8f059c7d0c0124e5b8a7bff84fd}{see Etherscan link}) as a representative case supporting \textbf{Obs1} and \textbf{Obs3} in \S\ref{sec:observations}.
\baselinename\ tends to focus on low-level suspicious patterns while overlooking the high-level semantic intent of contract behavior, often leading to overly conservative judgments and false positives. In this example, it incorrectly labels a benign contract as adversarial because it ``contains several unusual patterns and complex logic that are often seen in adversarial contracts'', even though these patterns represent legitimate access-control and compliance mechanisms. This misinterpretation reflects a lack of contextual reasoning; as noted in \textbf{Obs3}, \baselinename\ also defaults to adversarial predictions in ambiguous cases involving non-standard interactions.
In contrast, \name\ performs multi-view semantic analysis that integrates diverse behavioral signals. The six prompt scores are summarized in Table~X, and after entropy-based fusion, the aggregated scores for \texttt{adversarial}, \texttt{suspicion}, \texttt{uncertain}, and \texttt{benign} are 0.1211, 0.2472, 0.3225, and 0.3092, respectively. The combined adversarial score (0.37) leads to the correct benign classification.

Overall, \baselinename’s pattern-driven reasoning causes over-detection, whereas \name’s holistic and confidence-aware fusion yields more accurate and interpretable outcomes, confirming the value of multi-level semantic reasoning for robust adversarial contract detection.

\begin{figure}[t]
\begin{ResultBox}
\scriptsize\ttfamily

\blue{\textbf{Analysis of \baselinename:}}
\begin{itemize}
    \item The contract \textbf{contains several unusual patterns and complex logic that are often seen in adversarial contracts} designed to exploit or bypass security measures. 

    \item There are \textbf{numerous checks and updates} of state variables that appear convoluted and overlapping, which might be intended to obfuscate the true purpose of the contract. 

    \item There are repetitive and redundant operations around balance adjustments and transfer mechanisms, including states like `stor\_19\_21\_22`, whose intended roles and necessity are not clear, raising concerns about hidden or malicious functionality. 


    \item The contract involves frequent interaction with an external contract, as seen from multiple external calls, which could be exploited if not properly managed or secured. 

    \item Lack of clear and straightforward logic, with many nested conditions and branches, often makes it easier for an adversary to insert malicious logic without immediate detection. 

    \item The presence of operations that transfer zero wei or call external functions with seemingly unnecessary actions could be a technique to cloak malicious payloads or test system constraints.

    \item ...

    \item The contract performs complex and numerous checks on member access, transaction amounts, and wallet sizes for transfers, which could be indicative of attempts to deliberately manipulate such transactions for unauthorized gains. 

    \item The contract utilizes low-level operations (`ext\_code.size`, external calls) with conditions that may attempt to manipulate contract behavior unexpectedly, showing suspicious activity aligned with adversarial intent. 
\end{itemize}

\blue{\textbf{Final Prediction:}} 
\vspace{-5pt}

Adversarial \hspace*{4cm} \textbf{\textcolor{red}{\Huge\ding{55}}}

\end{ResultBox}
\caption{\baselinename detection results for the benign contract in Appendix \ref{sec:ablation_case}.}
\label{fig:ablation_case_prompt4}
\end{figure}

\begin{table}[htbp]
\centering
\begin{tabular}{l c c c c c c c c}
\toprule
\textbf{Prompt} & \textbf{G1} & \textbf{P1} & \textbf{G2} & \textbf{P2} & \textbf{G3} & \textbf{P3} & \textbf{G4} & \textbf{P4} \\
\midrule
prompt\_g\_normal      & D & 50 & C & 30 & B & 15 & A & 5 \\
prompt\_s\_normal      & B & 50 & C & 30 & A & 15 & D & 5 \\
prompt\_g\_mislead\_ad & C & 40 & D & 30 & B & 20 & A & 10 \\
prompt\_g\_mislead\_be & D & 70 & C & 20 & B & 8  & A & 2 \\
prompt\_s\_mislead\_ad & A & 70 & B & 20 & C & 8  & D & 2 \\
prompt\_s\_mislead\_be & D & 70 & C & 20 & B & 8  & A & 2 \\
\bottomrule
\end{tabular}
\caption{Predictions for the contract in Appendix \ref{sec:ablation_case}.}
\label{tab:case_study_prompt_result}
\end{table}

\end{document}